\documentclass[usenatbib]{aa}  

\usepackage{graphicx}
\usepackage{subcaption}
\usepackage{txfonts}
\usepackage{gensymb}
\usepackage{siunitx}
\usepackage[pdfencoding=auto,psdextra]{hyperref}
\hypersetup{
    colorlinks=true,
    linkcolor=blue,
    filecolor=magenta,      
    urlcolor=blue,
    citecolor=blue
}
\usepackage{xcolor}

\begin{document}

   \title{Cosmic Tunnels and the Integrated Sachs-Wolfe effect}

   \author{C. T. Davies
          \inst{1}
          M. Klein
          \inst{1}
          A. Fumagalli
          \inst{1,2} 
          and
          J.J. Mohr
          \inst{1}
          }

   \institute{$^{1}$ University Observatory, Faculty of Physics, Ludwig-Maximilians-Universität, Scheinerstr. 1, 81679, Munich, Germany\\
   $^{2}$ INAF-Osservatorio Astronomico di Trieste, Via G. B. Tiepolo 11, 34143 Trieste, Italy\\
              \email{C.Davies@physik.uni-muenchen.de}
              }

   \date{Received XXX; accepted YYY}

 
  \abstract
   {Cosmic voids, vast underdensities in the large-scale structure, offer unique sensitivity to cosmological parameters. However, traditional 3D galaxy-based void finding is limited by many factors, including uncertainties in the galaxy–halo connection and distortions from redshift errors. Using alternative tracers and new 2D void definitions can alleviate these limitations and be tailored to maximise the signal for specific observables.}
   {Here, we introduce Cosmic Tunnels, a new class of 2D void-like objects traced by galaxy clusters, corresponding to large underdense lines of sight.}
   {We identify Cosmic Tunnels by applying the tunnel algorithm to the RASS-MCMF and ACT-MCMF cluster catalogues. We validate their void-like nature by measuring the cross-correlation of Cosmic Tunnels with galaxy density contrast maps from the DESI Legacy Survey (measured at 63$\sigma$ significance), galaxy weak-lensing maps from DES Y3 (31$\sigma$), and CMB weak-lensing maps from ACT DR6 (15$\sigma$), all of which show underdense interiors enclosed by compensation ridges, consistent with 3D galaxy voids, measured at high statistical significance. We also show that the lensing profiles fit the universal HSW void profile, further validating their void-like nature.}
   {We cross-correlate the Cosmic Tunnels with Planck CMB temperature maps to measure the ISW signal. Using the ACT-MCMF Cosmic Tunnels, we achieve a 3.6$\sigma$ ISW detection, one of the highest significances ever reported from a single tracer catalogue. These results confirm that Cosmic Tunnels are robust underdense structures and demonstrate their potential as a new tool for cosmological analyses. Finally, we report a tentative detection of a sign flip in the ISW signal at very low redshift (z<0.03), consistent with previous studies, challenging the standard $\Lambda\rm{CDM}$ paradigm.}
   {}

   \keywords{
               }

   \maketitle
%

\section{Introduction}

The large-scale structure (LSS) of the Universe contains key information about the growth of cosmic structures and cosmology. It arises from the anisotropic gravitational collapse of initially small density perturbations in the early Universe \citep{Zel'dovich1970}. The resulting structure, also known as the cosmic web, can be broadly classified into four morphological components: knots, filaments, walls, and voids \citep{Kirshner1981, Davis1985, Bond1996}. Knots represent the largest overdensities in the Universe, typically located at the intersections of filaments, which themselves form from overlapping walls. These structures enclose vast underdense regions known as cosmic voids, which dominate the volume of the Universe \citep[e.g.][]{Padilla2005, Platen2007, Cautun2013}.

The cosmic web can be characterised with various summary statistics of these components. For overdense regions, this includes the halo mass function \citep[HMF,][]{Tinker2008} and the galaxy two-point correlation function \citep{Totsuji1969}. For underdense regions, analogous statistics such as the void size function (VSF), the void correlation function exist \citep{Hamaus2014, Pisani2015, Sutter2014, Contarini2019, Fang2019}. These summary statistics are sensitive to non-Gaussian features of the matter distribution that are not optimally captured by traditional two-point statistics such as the matter correlation function and the matter power spectrum \citep{Colberg2008}. In particular, both the HMF and VSF encode additional information about the non-linear regime of structure formation and have therefore emerged as powerful tools for cosmological inference \citep[e.g.][]{Hamaus2016,Bocquet2024}.

Despite their potential, void-based cosmological constraints have historically been weaker than those obtained from galaxy clusters. This is primarily due to limitations in the tracer populations and void identification methods. Typically, galaxies are used as tracers of the matter field for void identification. However, the galaxy distribution is sparse, and uncertainties remain in the galaxy-halo connection. Modelling the galaxy-halo connection introduces additional parameters that are degenerate with the cosmological parameters of interest (such as in the halo occupation distribution (HOD) model \citep{Zheng2005}), and this degrades constraints. Additionally, photometric redshift uncertainties distort galaxy positions along the line of sight, reducing the reliability of void identification \citep{Mao2017}. Even with accurate redshifts, many void-finding algorithms, such as those based on density minima or watershed techniques, produce spurious voids. This requires additional cleaning steps that inadvertently remove some true voids and reduce the available signal \citep{Contarini2019}.

Alternative tracers for void finding have been proposed to circumvent these issues. These include optically selected galaxy clusters \citep{Pollina2019}, the Lyman-alpha forest \citep{Krolewski2018}, and peaks in the weak lensing field (WL) \citep{Davies2018, Davies2019b, Davies2020, Davies2020b, Maggiore2025}. Notably, voids identified using 2D data, such as WL maps or the projected galaxy distribution, have demonstrated consistency with underdensities in the 3D matter field \citep[][e.g.]{Gruen2015, Jeffrey2021}. Because the line of sight information is collapsed in 2D voids, redshift uncertainties become significantly less impactful.

Void identification methods vary widely depending not only on the tracer population, but also on the algorithm employed. Numerous studies have compared the efficacy of different void definitions, highlighting significant differences in the resulting void populations and their inferred properties \citep{Colberg2008, Cautun2018, Paillas2019, Davies2020b}. For instance, comparisons between watershed voids, spherical void finders, and 2D voids such as troughs and tunnels, show that the void abundances and profiles can vary dramatically depending on the methodology. These discrepancies emphasise that the void statistics are not uniquely defined but rather depend on the adopted definition. Although this ambiguity can pose challenges for theoretical modelling, with careful design, void-finding algorithms can be tailored to maximise the signal-to-noise ratio (SNR) of a chosen summary statistic while minimising systematic uncertainties. Such a pipeline then requires a robust forward model from simulations in place of a theoretical model. 

Recent studies have demonstrated the utility of cosmic voids as cosmological probes. Voids have been employed to test and constrain $\Lambda$CDM parameters via the void abundance \cite{Contarini2023}, the Alcock–Paczynski (AP) test \citep{Lavaux2012, Hamaus2016}, constrain modified gravity models \citep{Barreira2015, Achitouv2016}, probe the sum of neutrino masses \citep{Massara2015, Kreisch2019}, and detect the Integrated Sachs–Wolfe (ISW) effect \citep{Granett2008, Kovacs2018, Hang2021}. These diverse applications highlight the growing maturity of void science as a tool for cosmological inference.

In this work, we propose a new method for identifying cosmic voids using galaxy cluster positions to define 2D voids via the tunnel algorithm \citep{Cautun2018}. Galaxy clusters detected through Sunyaev-Zel’dovich (SZ) and X-ray surveys provide a high-bias tracer population with well-calibrated masses and reduced galaxy-halo connection uncertainties \citep{Bocquet2024}. However, their low number density is limiting for conventional 3D void finding. Instead, we focus on 2D void identification, which is well-suited to low-density tracers and less sensitive to redshift uncertainties. This approach leverages the strengths of cluster datasets while mitigating the limitations in galaxy-based void studies. The objects identified in this way correspond to large underdense lines of sight. 

This paper is outlined as follows. In Sec. \ref{sec:tracer data} we describe the X-ray and SZ selected galaxy cluster catalogues from \cite{RASSMCMF,klein2024actdr5} that are used in this work. These galaxy cluster catalogues are used to identify the 2D voids via the tunnel algorithm in Sec. \ref{sec:tunnel finder}, which we name Cosmic Tunnels. In Sec. \ref{sec:Cosmic Tunnel abundance} we investigate the size distribution of the Cosmic Tunnels and interpret the results in terms of the tracers population. Then, to validate that Cosmic Tunnels correspond to true underdense lines-of-sight, we perform a series of cross correlations. We first cross-correlate the Cosmic Tunnels with the WL field in Sec. \ref{sec:lensing profiles} using the galaxy WL data from DES \citep{Jeffrey2021}, and CMB WL data from ACT \citep{Madhavacheril2024}. We find that the lensing cross correlations verify that the Cosmic Tunnels correspond to underdense lines of sight, both for galaxy and CMB WL. Remarkably, we also find that the lensing profiles of the Cosmic Tunnels can be fit by the universal HSW void profile \citep{Hamaus2014}, providing even further evidence for the void-like nature of Cosmic Tunnels. In Sec. \ref{sec:gal profiles} we investigate the Cosmic Tunnel-galaxy cross correlation with galaxy catalogues from the Legacy Survery \citep{Legacysurveys19}. We find that Cosmic Tunnels enclose statistically significant underdensities in the galaxy population, with a compensation ridge at the void boundary, as expected for void-like objects. Finally, in Sec. \ref{sec:isw} we investigate the potential for Cosmic Tunnels to measure the Integrated Sachs-Wolfe effect (ISW). We achieve this through a cross correlation with the CMB temperature anisotropy map from Planck \citep{Planck2018}, and report a 3.6$\sigma$ detection of the ISW signal with Cosmic Tunnels. Finally in Sec. \ref{sec:lowzisw} we report the detection of a sign flip in the ISW signal at low redshift, consistent with recent studies \citep{Hansen2025}, although we caution that the low redshift ISW measurements have low statistical significance.

\section{Data}

\begin{figure*}
    \centering
    \includegraphics[width=2\columnwidth]{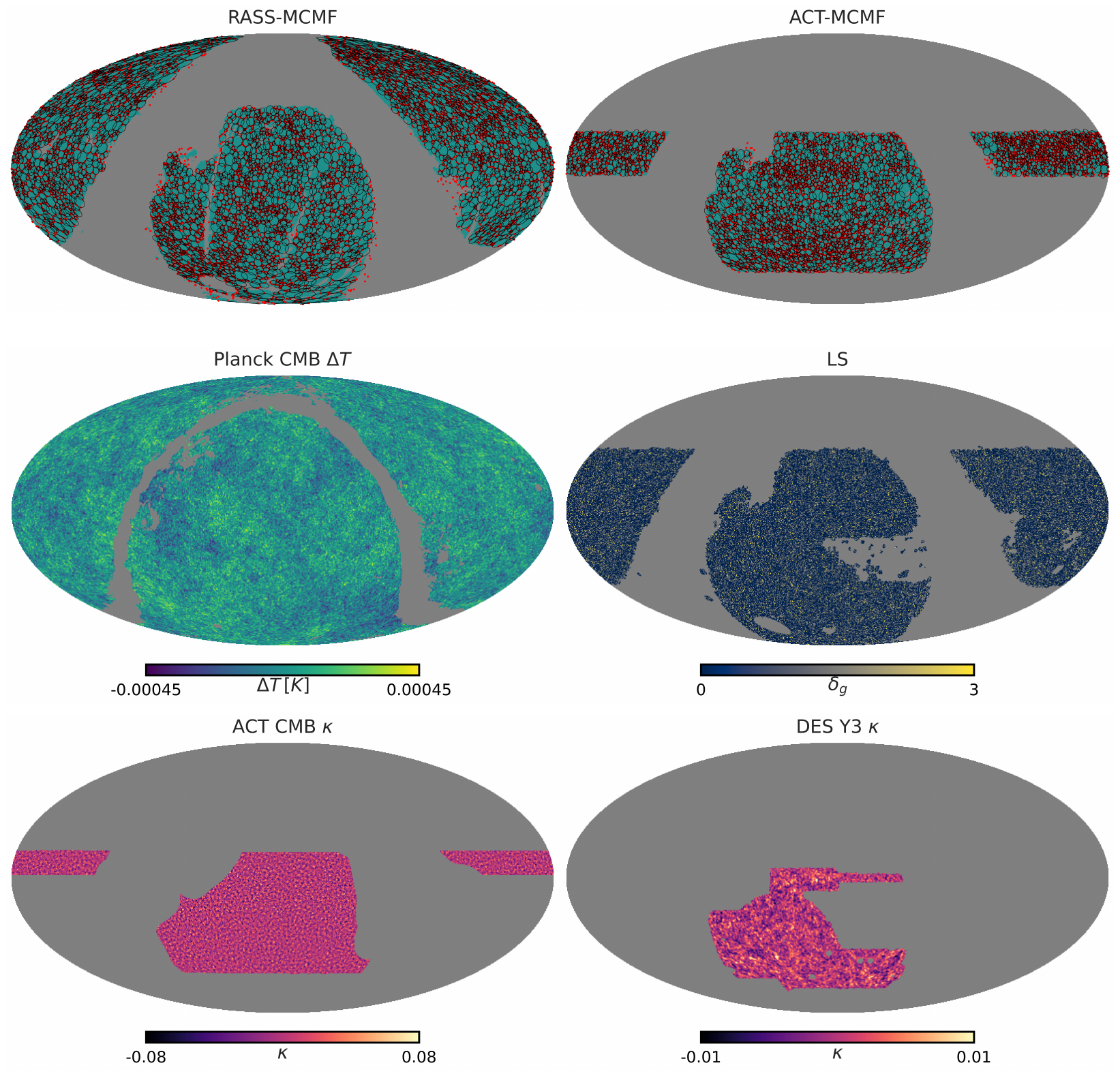}
    \caption{The observational used in this analysis, the shaded grey regions indicate the survey mask associated with the data in each panel. Colour bars indicate the continuous variable with its range and units used in each panel. The top two panels show the Cosmic Tunnels (black) identified in RASS-MCMF (left) and ACT-MCMF (right), using all clusters in the catalogues, which are shown by the red points. The second row shows the Planck CMB temperature anisotropy map (right), and the Legacy Survey projected galaxy overdensity field for $z\in[0,0.5]$(left). The bottom row shows the ACT CMB WL convergence map (left) and the DES-Y3 galaxy WL convergence map (right).}
    \label{fig:cross maps}
\end{figure*}

In this section, we describe the data used in this analysis, which can be split into two types, tracer data and cross-correlation data. Tracer data are the data that is combined with a void finding algorithm (outlined in Sec. \ref{sec:tunnel finder}) to identify Cosmic Tunnels. Cross-correlation data are additional data that are used to validate the void-like nature of Cosmic Tunnels in multiple regimes. The cross-correlation data can be further split into three categories, WL data, galaxy data, and CMB temperature data. The data used in this analysis are plotted in Fig. \ref{fig:cross maps}, which is referenced throughout the text, and provides a clear visual picture of the analysis carried out in this work. It is important to consider a range of cross-correlation data to robustly and independently validate the void-like nature of Cosmic Tunnels without being limited by the systematics associated with a single type of dataset. 

The tracers used in this work are galaxy clusters, which are taken from the X-ray-selected cluster catalogue, RASS-MCMF \citep{RASSMCMF}, and the SZ-selected cluster catalogue from ACT-MCMF \citep{klein2024actdr5}. 

The cross-correlation data are outlined as follows. For the WL data, we use the galaxy-lensing convergence maps from DES Y3 \citep{Jeffrey2021}, and CMB-lensing convergence maps from ACT \citep{Madhavacheril2024}. For the galaxy distribution, we use data from the legacy survey \citep{Legacysurveys19}. Finally, for the CMB temperature data, we use the temperature anisotropy map from Planck \citep{Planck2018}.

\subsection{Tracer data}\label{sec:tracer data}
This section outlines the tracer data employed in this analysis, which is the data that are provided to the void finder (Tunnel algorithm), to identify Cosmic Tunnels. Here the tracer data correspond to galaxy cluster positions from the RASS-MCMF and ACT-MCMF catalogues. 

\begin{figure}
    \centering
    \includegraphics[width=\columnwidth]{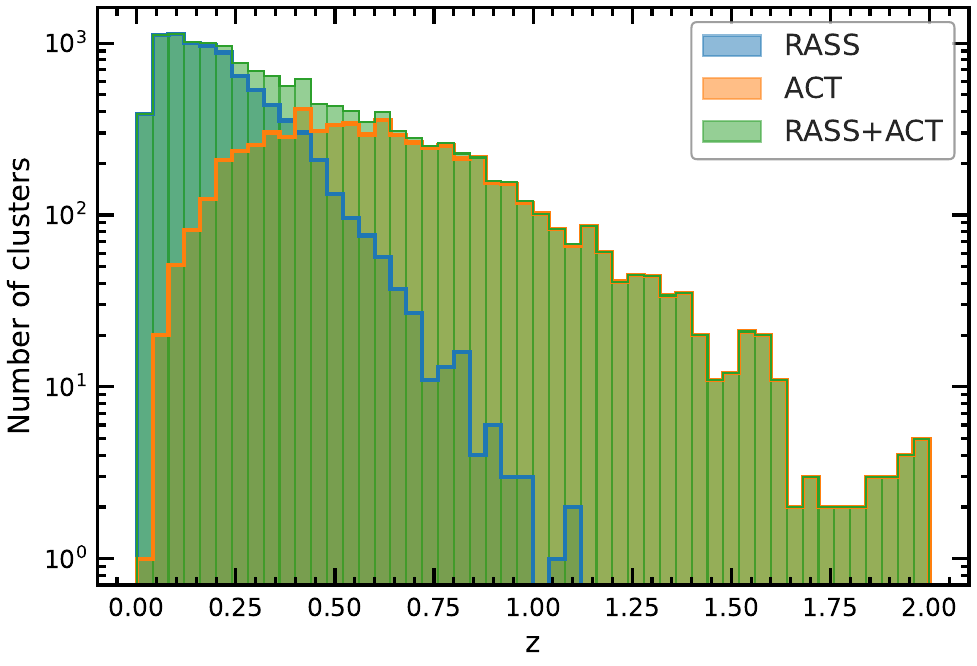}
    \caption{The redshift distribution of the galaxy cluster catalogues used in this analysis. The blue and orange curves show the data for the RASS and ACT MCMF catalogues respectively. The orange curve shows the combination of the two catalogues, where duplicate clusters present in both catalogues have been removed.}
    \label{fig:cluster catalogues}
\end{figure}

\subsubsection{RASS-MCMF cluster catalogue}

The RASS-MCMF cluster catalogue \citep{RASSMCMF} consists of \num{8449} X-ray-selected galaxy clusters over $\sim$\num{25000} deg$^2$ of good extragalactic sky. The X-ray selection is based on the 2RXS catalogue \citep{Boller2RXS} of X-ray sources from the ROSAT All Sky Survey \citep{Truemper93}. The cluster catalogue is constructed from the X-ray source list using the multi-component matched filter (MCMF) method \citep{2018MNRAS.474.3324K,MARDY3} utilising optical imaging and photometric data from the tenth data release (DR10) of the DESI Legacy Survey \citep{Legacysurveys19}. The RASS-MCMF cluster selection only depends on X-ray detection likelihood and optical richness as a function of redshift. The baseline catalogue has an expected purity of 90\% with respect to X-ray selected galaxy clusters. The footprint of the RASS-MCMF survey is shown in Fig.~\ref{fig:cross maps}.

\subsubsection{ACT-MCMF cluster catalogue}
The ACT-DR5 MCMF catalogue \citep{klein2024actdr5} covers $\sim$\num{13000} deg$^2$ of the equatorial and southern sky and consists of \num{6247} SZ-selected and optically confirmed galaxy clusters. The source catalogue is based on data from the fifth data release of the Atacama Cosmology Telescope \citep{ACTtel,ACTDR5data}. The source detection pipeline is the same as that used for the original ACT-DR5 catalogue \citep{ACTDR5} with only slightly different settings. Consistent with RASS-MCMF, the construction of the optically confirmed cluster catalogue is based on the MCMF method and uses the DESI Legacy Survey DR10 dataset. Similarly to previous work on SZ-selected clusters from the South Pole Telescope \citep{SPTSZMCMF,2024OJAp....7E..13B}, an additional MCMF-based confirmation step was applied using infrared data from the WISE satellite \citep{Wright_2010,UnWISE} to confirm galaxy clusters at high redshifts. This additional step is needed because the SZ-selection is greatly redshift independent, allowing for the detection of galaxy clusters at high redshift in contrast to X-ray-based samples, which drop significantly in sensitivity at high redshifts.

Fig. \ref{fig:cluster catalogues} shows the cluster redshift distributions for RASS-MCMF (blue) and ACT-MCMF (orange). The total number of clusters used here is 8449 for RASS-MCMF and 6237 for ACT-MCMF, both with a purity of $90\%$. The plot also shows the redshift distribution of the combination of the two catalogues, labelled RASS+ACT (green), where duplicate clusters are removed by searching for objects present in both catalogues within five arcminutes of each other. With this approach, we find that there are 1064 duplicate clusters between the two catalogues, leading to a total number of 13,622 clusters in the RASS+ACT catalogue. The creation of the joint RASS+ACT catalogue will allow for additional investigations with Cosmic Tunnels that benefit from the increased tracer density of the combined catalogue. We note that the increased tracer density is only present in the regions of the sky where the two catalogues overlap. For the cross correlations performed with the RASS+ACT catalogue, unless otherwise stated (in Sec. \ref{sec:isw}), we include the entire sky area covered by the two catalogues to increase the statistical precision. 

The figure illustrates that the number of clusters in the RASS X-ray-selected sample peaks at low redshift ($\approx 0.15$), with a maximum redshift of $z=1.2$. For the ACT SZ-selected sample, the distribution peaks at an intermediate redshift of $z=0.5$, with a maximum redshift of $z = 2$. This quantitatively illustrates how the two cluster catalogues probe different regimes of the LSS, which will naturally lead to differences in the Cosmic Tunnel statistics measured from the two samples. The figure also illustrates that the joint RASS+ACT catalogue is sourced mostly by the RASS catalogue below $z=0.25$, and dominated by ACT above $z=0.6$, where the intermediate range $z\in[0.25,0.6]$ is where the RASS+ACT catalogue receives roughly equal contributions from both of the original catalogues.  

\subsection{Cross correlation data}\label{sec:cross correlation data}
This section outlines the data used to perform cross-correlations with the Cosmic Tunnels identified from the tracer data. The cross correlations are used to validate the void-like nature of Cosmic Tunnels in various regimes.  

\subsubsection{DES Y3 mass map}
To measure the galaxy weak-lensing signal from Cosmic Tunnels, we use the DES Y3 convergence map from \cite{Jeffrey2021}. The convergence, $\kappa$, at a given position on the sky, $\pmb{\theta}$, from a source at a given comoving distance, $\chi$, from the observer is given by the following expression

\begin{equation}
    \kappa(\pmb{\theta},\chi) = \frac{3H_0^2\Omega_{\rm{m}}}{2c^2}\int_0^{\chi}\frac{\chi - \chi'}{\chi} \chi' \frac{\delta(\chi'\pmb{\theta},\chi')}{a(\chi')} {\rm d}\chi'.
    \label{eq:conv source}
\end{equation}
Here $H_0$ is the Hubble constant at $z=0$, $c$ is the speed of light, and $\Omega_{\rm m}$ is the matter density parameter at $z=0$, $a$ is the scale factor of the universe at $z=z(\chi)$, and $\delta$ is the overdensity parameter at $z=z(\chi)$. This expression shows that the convergence signal is a weighted projection of the matter field, where the weighting can be attributed to the lensing kernel $\chi'(\chi-\chi')/\chi$, which is a parabola that peaks at $\chi/2$.

The source galaxies used to measure the lensing signal are not observed in a fixed source plane, but span a range of distances from the observer. Therefore, the observed $\kappa$ signal in Eq. \eqref{eq:conv source} must be weighted by the observed source galaxy distribution $n(\chi)$, which gives $\kappa(\pmb{\theta})$ \citep[see, e.g.,][for details]{Kilbinger2015}, and is defined as follows

\begin{equation}
    \kappa(\pmb{\theta}) = \int_0^{\chi_{\rm H}} n(\chi') \kappa(\pmb{\theta},\chi') {\rm d}\chi' \, ,
    \label{eq:kappa obs}
\end{equation}
where $\chi_{\rm H}$ is the comoving distance to the horizon. The exact form of $n(\chi)$ for the corresponding $\kappa$ map is presented in \cite{Jeffrey2021}. For the DES Y3 data, the $n(\chi)$ distribution peaks at around $z \approx 0.6$, with a maximum redshift extent of $z \approx 1.5$. We note that the $n(\chi)$ of the source galaxy sample can be split into tomographic bins, to allow for variations in the source redshift which probes structure at different times. However, in this work we use the full $n(\chi)$ which corresponds to the non-tomographic case (i.e. the sum of the tomographic bins), as we are investigating 2D cosmic tunnels, and are therefore interested in the whole matter distribution along the line of sight. This formalism illustrates that the convergence signal corresponds to the total matter density integrated along the line of sight, weighted by the lensing kernel (which peaks at half the comoving distance from the observer to the source), which makes the DES Y3 $\kappa$ map an excellent data-set for testing the void-like nature of Cosmic tunnels and hence verifying that they correspond to underdense lines-of-sight.

The DES-Y3 $\kappa$ map is shown in the bottom right panel of Fig. \ref{fig:cross maps}, with the corresponding colour bar, which shows the $\kappa$ values throughout the map. The DES-Y3 $\kappa$ map has a total sky coverage of 4132 deg$^2$ and a source density of 5.59 gal arcmin$^{-2}$. The map is smoothed with a Gaussian filter with width given by a standard deviation of 10 arcminutes. We note that this smoothing scale is of a comparable size to the smallest voids plotted in Fig. \ref{fig:cosmic tunnel abundance}, which is discussed in Sec. \ref{sec:Cosmic Tunnel abundance}. 

\subsubsection{ACT DR6 mass map}
To measure the CMB WL signal from Cosmic Tunnels, we use the ACT DR6 lensing convergence map from \cite{Madhavacheril2024}. Physically, the convergence signal in the CMB lensing map can be described by Eq. \ref{eq:kappa obs}, where the source redshift distribution corresponds (to good approximation) to a delta function at the redshift of recombination. As a result, the CMB lensing kernel peaks at $z\approx2$ \citep{Madhavacheril2024}, making it sensitive to higher redshifts than galaxy WL, which probes lower-redshift structure via the shapes of background galaxies.

The CMB lensing signal is reconstructed using a quadratic estimator that combines pairs of CMB temperature and polarisation fields (e.g., TT, TE, EE, EB) in \cite{Madhavacheril2024} to produce $\kappa_{\ell m}$'s, which is the data used in this work. Before transforming to real space, we remove modes outside the range $100<\ell<2090$. The upper cut-off reflects a balance between resolution and noise: beyond this point, the lensing reconstruction becomes increasingly noise-dominated due to instrumental limitations and filtering in the ACT DR6 pipeline. Additionally, the angular scales corresponding to these higher multipoles are much smaller than the voids in our sample, and do not contribute meaningful structure to the stacked lensing signal. The lower cut removes large-scale modes ($\ell < 100;\theta \geq 2 ^{\circ}$) that would otherwise introduce long-range correlated noise across the map but would be poorly resampled by our covariance estimator (described in 
Appendix. \ref{sec:SNR}). The reconstruction noise in the CMB lensing maps is not white; it varies across angular scales because of the scale-dependent filtering of the primary CMB anisotropies, instrumental noise, and the effects of beam smoothing and masking. In contrast, galaxy lensing noise can typically be approximated as white in harmonic space, since it is dominated by the uncorrelated intrinsic shape noise of individual galaxies.

The ACT $\kappa$ map is shown in the bottom left panel of Fig. \ref{fig:cross maps}, with the corresponding colour bar, which shows the $\kappa$ values throughout the map. The ACT $\kappa$ map has a total sky coverage of 9400 deg$^2$. We note that the highest $\ell$ included in the map reconstruction corresponds to an angular extent of roughly 5 arcminutes, which is indicative of the smallest physical scales present in the ACT $\kappa$ map. This is half the smoothing employed in the DES Y3 $\kappa$ map, and again, this corresponds roughly to the smallest voids plotted in Fig. \ref{fig:cosmic tunnel abundance}, which is discussed in Sec. \ref{sec:Cosmic Tunnel abundance}.

\subsubsection{Legacy survey galaxies}

The projected galaxy-density field we use to validate Cosmic Tunnels is built from the public DESI Legacy Imaging Surveys (hereafter Legacy Surveys). The Legacy Surveys comprise three wide-area optical programmes—DECaLS (Dark Energy Camera Legacy Survey), BASS (Beijing–Arizona Sky Survey) and MzLS (Mayall $z$-band Legacy Survey)—supplemented by forced photometry from the unWISE re-processing of WISE/NEOWISE mid-IR images. While MzLS and BASS predominantly cover the northern sky above declination $30\ \mathrm{deg}$ in $g, r$ and $z$-band respectively, DECALS covers the sky below that declination in all three bands observed with DECam \citep{Flaugher15} at the 4m Blanco telescope. In subsequent data releases additional archival data from other DECam-based surveys such as DES \citep{DES2016}, DELVE \citep{DELVE}, and DeROSITAS \citep{DEROSITAS} were included, increasing the sky coverage in the southern hemisphere, culminating in a coverage of $\approx25,0000\ \mathrm{deg}^2$ of the extragalactic sky \citep{RASSMCMF}.


For this analysis we use DR10 of the Legacy Surveys, limited to the sky observed by DECam with $g, r, z$-band coverage and available photometric redshifts. The footprint is shown in the middle right panel of Fig. \ref{fig:cross maps}. The typical $g$, $r$, $z$-band median $5\sigma$ depths are $g\simeq24.7$, $r\simeq24.2$ and $z\simeq23.3$ (AB) outside the DES footprint and $g\simeq25.3$, $g\simeq25.0$, and $g\simeq23.9$ inside. 

We use the DR10 version of the photo-z catalogue \citep{Zou2023} to impose a redshhift cut of $0.05< z < 0.5$ and select galaxies brighter than 21.3 mag in $z$-band, corresponding to a stellar mass of $\approx 10^10\ \mathrm{M}_\odot$. We further reduce stellar contamination by excluding all sources with the morphology type "PSF" in the catalogue. These cuts result in a clean and complete galaxy catlogue, even in the shallower regions of the survey footprint.
M
We project the galaxy catalogue onto a \textsc{HEALPix} map with $N_{\mathrm{side}}=4096$ (pixel size $\sim0.9'$) and calculate the fractional galaxy overdensity after applying the survey mask as
\begin{equation}
\delta_g(\mathbf{\theta}) \equiv \frac{N(\hat{\mathbf{\theta}}) - \bar{N}}{\bar{N}} \, ,
\label{eq:galdencon}
\end{equation}
where $N(\pmb{\theta})$ is the number of galaxies in the \textsc{HEALPix} pixel located at $\pmb{\theta}$, and $\bar{N}$ is the mean number of galaxies in all \textsc{HEALPix} pixels. This two-dimensional density contrast map is then used to measure the radial galaxy-density profiles around every Cosmic Tunnel.

The Legacy Surveys galaxy density contrast map is shown in the middle right panel of Fig. \ref{fig:cross maps}, with the corresponding colour bar, which shows the $\delta_g$ values throughout the map. The map has a total sky coverage of $\approx16000$ deg$^2$, with an NSIDE of 4096, which corresponds to 0.9 arcminutes. This is much smaller than the scales available in the WL maps, and smaller than the smallest Cosmic Tunnels investigated in this work. 

\subsubsection{Planck CMB map}

After using WL and galaxy density data to validate that Cosmic Tunnels are underdense, we also perform a cross-correlation between Cosmic Tunnels and the CMB temperature anisotropy signal from the Planck 2018 temperature map \citep{Planck2018}. This allows us to measure the corresponding Integrated Sachs-Wolfe (ISW) signal from the CMB.

We use the foreground-cleaned Commander CMB temperature map from the Planck Legacy Archive, along with the associated survey mask to exclude regions contaminated by Galactic emission or strong point sources. Given that the CMB map will be used to measure the ISW effect through cross correlation with the Cosmic Tunnels, which is a low- and cosmic-variance limited signal, it is important to also construct an accurate null measurement to allow for robust signal-to-noise measurements. To this end, we employ the FFP10 simulations \cite{Planck2020}, which provide 1000 monte-carlo realisations of the CMB sky, including instrumental and systematic effects processed through the same pipelines as the data.

The Planck CMB temperature anisotropy map is shown in the middle left panel of Fig. \ref{fig:cross maps}, with the corresponding colour bar, which shows the $\Delta T$ values throughout the map. The map has a total sky coverage of 32,200 deg$^2$, with an NSIDE of 2048, which corresponds to a pixel resolution of 1.7 arcminutes, however the beam width is 5 arcminutes, which sets the limit for the physically interpretable resolution. 


\section{Methodology}
In this section, we outline the methods used in this analysis. We first present the Tunnel algorithm which is used to identify tunnels (2D voids on the projected sky) from a set of discrete tracers. This is followed by a discussion on the HSW profile, a universal void profile known to provide strong characterisation of underdensities. Finally, we present the methodology that we employ to evaluate the signal-to-noise ratios of all of the cross correlations presented in this study. 

\subsection{Tunnel void finder}\label{sec:tunnel finder}
The tunnel algorithm \citep{Cautun2018} is a 2D void finder that identifies the largest circles that are empty of tracers \footnote{The tunnel algorithm used in this work is publicly available at \href{https://github.com/chrisdavies234/tunnel_finder}{https://github.com/chrisdavies234/tunnel\_finder}}. This is achieved by first constructing a Delaunay triangulation out of a set of discrete points \footnote{Note that the Delaunay triangulation corresponds to the dual graph of the Voronoi diagram. The Voronoi diagram is a popular method for estimating the underlying density field sampled by a set of tracers, and commonly employed in other void finding algorithms.}. Each cell in the Delaunay triangulation is defined as a triangle whose vertices correspond to a tracer, where each triangle does not enclose any tracers. These cells are then used to define a circumcircle whose circumference intersects the three vertices of its corresponding triangle, and hence the circumcircles also contain no tracers. These circumcircles correspond to the tunnels (2D voids) identified by the algorithm. 

In this work, we use the positions of the galaxy clusters as the tracers for the tunnel algorithm outlined above. This allows us to identify large circles on the sky that correspond to extended lines of sight that do not enclose any (detected) galaxy clusters, which we call Cosmic Tunnels. 

To increase the robustness of our Cosmic Tunnel sample, we apply some additional cleaning criteria to the catalogue that is output by the tunnel algorithm. 

First, we remove circumcircles that correspond to Delaunay triangles whose minimum enclosed angle is less than 20 degrees. Such triangles are highly elongated along one direction, and thus have an area much smaller than their corresponding circumcircle, which is undesirable. 

Second, as an optional step in some cases, we remove circumcircles whose centres are enclosed within a larger circumcircle, which gives priority to identifying the largest objects, whilst simultaneously reducing the overlap of the circumcircles and hence minimises the duplicate information throughout the catalogue. We note that the no-overlap criteria is only applied to Cosmic Tunnel catalogue used in the ISW measurements in Sec. \ref{sec:isw}, where careful consideration must be taken to minimise spurious correlations and features that may falsely inflate the ISW detection significance. For other cross-correlations, such as with the WL and galaxy density contrast fields, we allow the Cosmic Tunnel catalogues to include overlapping objects. These signals have an intrinsically higher SNR compared to the ISW measurements, and so it is safe to use this approach to minimise the bin-to-bin scatter. However, it will also be important to account for the bin-to-bin correlation this induces, which can be captured through measuring the corresponding covariance matrices \cite{Davies2019b}.  

Since the tunnel algorithm is applied to tracers from survey data, we must also account for the survey mask. Throughout this work, we remove circumcircles whose masked fraction of the enclosed area is above a given value, $f_{\rm mask}$. We have tested a range of values of $f_{\rm mask}$, and find qualitatively similar results for all cases, where the noise in various measurements with the Cosmic Tunnel increases as $f_{\rm mask}$ decreases, since this leads to fewer Cosmic Tunnels in the final catalogue. In this work, we use $f_{\rm mask} = 0.3$, which offers a good compromise between maximising the number of Cosmic Tunnels in the final catalogue, and minimising the overlap with the masked regions. 

We also note that the algorithm used here operates directly on the curved sky, performing the Delaunay triangulation directly on the surface of a unit sphere, which is the first time the Tunnel algorithm has been extended beyond the flat-sky approximation. 

When measuring the Cosmic Tunnel radial galaxy density profiles through the cross-correlation with the Legacy survey galaxies, we perform a local background subtraction to account for the survey inhomogeneity. For the background value corresponding to each void, we calculate the mean value of the legacy survey galaxy density contrast field at twice the void radius, within an annulus of the same width as the annuli used to bin the void profiles, and subtract this value from the void profile before stacking all voids. This forces the galaxy density contrast profile to cross zero at $r = 2 R_v$.

Finally, 
we perform a cross-correlation between the Cosmic Tunnels and the CMB temperature anisotropy, to measure the ISW effect from Cosmic Tunnels. A common approach to maximise the strength with which the ISW signal can be measured, is to apply to compensate top hat (CTH) filter to the CMB map when measuring the individual void profiles. To measure the CTH filtered temperature anisotropy profile $\Delta T_f(r/R_v)$, we compute the difference in the mean $\Delta T$ values within a disc of radius $r/R_v$, and an enclosing ring of equal area to the inner disc. This suppresses contributions to the signal from scales that are much smaller or larger than the filter scale. This step is computed for a range of filter scales, and stacked over all voids, returning a mean $\Delta T_f(r/R_v)$ for a given Cosmic Tunnel sample.

\subsection{HSW profile}\label{sec:HSW}
To test the void-like nature of Cosmic Tunnels, we fit the universal void profile - the HSW profile \citep{Hamaus2014} - to the WL profiles measured in this work. We note that the prescription for the HSW profile given in \cite{Hamaus2014} is expressed in terms of the void radial matter density profile. In this work, however, we instead express this in terms of a radial convergence profile, while keeping the same functional form as the original profile. We modify the notation, as the convergence corresponds to the projected total matter distribution along the line of sight. We also note that we do not bin the Cosmic Tunnels into radial-size bins before measuring the stacked radial profiles, as with typical HSW profile fits, and instead stack all voids by normalising the distance to the tunnel centre by the tunnel radius for each Cosmic Tunnel. This leads to an additional free parameter relative to the standard HSW profile, as the mean void size in a void size bin is typically used to fix one of the free parameters for a 3D HSW density profile. We make this choice to minimise scatter in the final measurement and remain as general as possible. We leave investigations of binning in void size to a future work. Due to the large range of Cosmic Tunnel sizes used when stacking over all voids, we instead leave the associated free parameter open. The form of the HSW profile we use in this work is then given by the following expression,

\begin{equation}
\kappa(r/R_v) = \kappa_c \frac{1 - \left( \frac{r/R_v}{r_i} \right)^{\alpha}}{1 + \left( \frac{r/R_v}{r_o} \right)^{\beta}} \, ,
\end{equation}
where $\kappa(r/R_v)$ is the WL convergence value at radius $r/R_v$, $\kappa_c$ controls the depth of the profile, $r_i$ is an inner scale radius that controls where the inner slope dominates, $r_o$ is the outer scale radius that controls where the outer slope dominates (normally the mean void size in the associated size bin in 3D void studies), and $\alpha$ and $\beta$ are shape parameters that control the inner and outer slope of the profile, respectively.

\section{Results}

In this section, we present the results of the analysis. First we present and discuss the direct outputs from the tunnel algorithm, the Cosmic Tunnel abundance, which is presented in Sec. \ref{sec:Cosmic Tunnel abundance}. Then we evaluate cross-correlations with WL and the galaxy distribution in Sec. \ref{sec:lensing profiles} and \ref{sec:gal profiles}, which is used to validate their void-like nature. Finally, we perform a cross-correlation with the Planck temperature anisotropy map, which we use to measure the ISW signal with Cosmic Tunnels. The detection significances of these cross-correlations are summarised in Table \ref{tab:snr}.

\begin{table}[h]
    \caption{Summary of the SNR's for various measurements with Cosmic Tunnels from the RASS, ACT, and RASS+ACT galaxy cluster catalogues. Cosmic Tunnel cross correlations include galaxy WL from DES Y3 \cite{}, CMB WL from ACT \cite{}, galaxy density contrast from the DESI legacy survey \cite{}, and CMB temperature anisotropy from Planck \cite{}.}
    \centering
    \textbf{SNR ($\sigma$)} \\[0.5em]  
    \begin{tabular}{lccc}
    \hline
     & \textbf{RASS} & \textbf{ACT} & \textbf{RASS+ACT} \\
    \hline
    \textbf{DES WL $\kappa(r/R_v)$} & 28.8 & 16.0 & 31.2  \\
    \textbf{ACT WL $\kappa(r/R_v)$} & 7.1 & 11.7 & 15.5 \\
    \textbf{DESI LS $\delta_g(r/R_v)$} & 49.5 & 42.9 & 63.2 \\
    \textbf{Planck $\Delta T_f$ ISW} & 3.1 & 3.6 & 2.7 \\
    \hline
    \end{tabular}
    \label{tab:snr}
\end{table}

\begin{figure}
    \centering
    \includegraphics[width=\columnwidth]{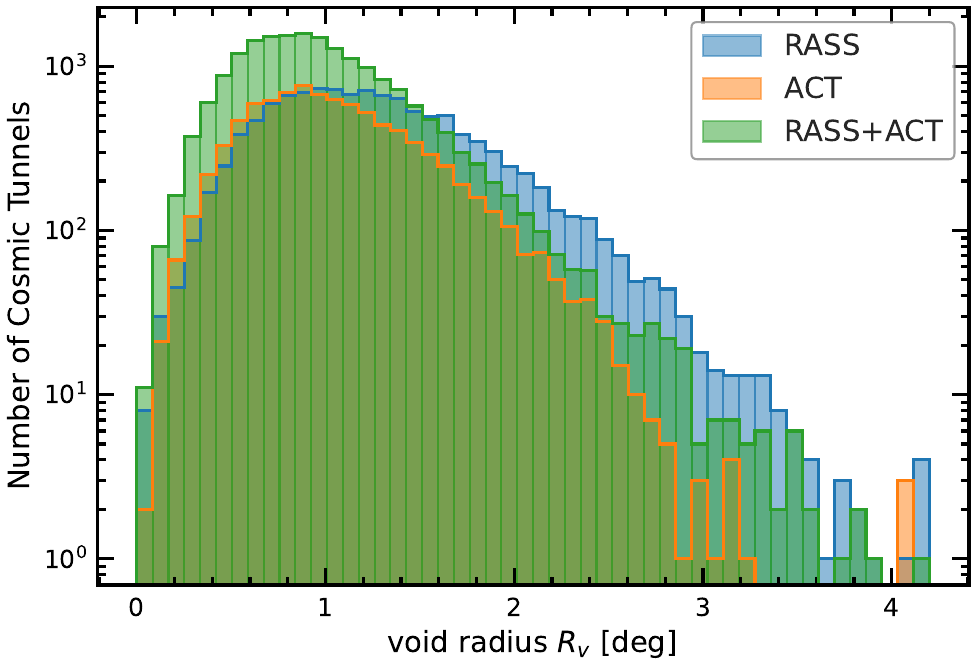}
    \caption{The cosmic tunnel abundance plotted as a function of the cosmic tunnel size $R_v$. The blue, orange, and green curves correspond to the cosmic tunnels identified in the RASS, ACT, and RASS+ACT MCMF galaxy cluster catalogues respectively. }
    \label{fig:cosmic tunnel abundance}
\end{figure}

\subsection{Cosmic Tunnel abundance}\label{sec:Cosmic Tunnel abundance}

We apply the Tunnel algorithm discussed in Sec. \ref{sec:tunnel finder} to the RASS, ACT, and RASS+ACT galaxy cluster catalogues presented in Sec. \ref{sec:tracer data} to generate the corresponding Cosmic Tunnel catalogues, and discuss the output size distribution. 
For the initial analysis, we do not apply any redshift cuts to the galaxy cluster catalogues before tunnel identification, and allow the tunnels to overlap. Each tunnel is then defined by a centre and a radius, where the boundary intersects with three galaxy clusters. 

The Cosmic Tunnel size distribution for the RASS, ACT, and RASS+ACT catalogues is presented in Fig. \ref{fig:cosmic tunnel abundance} and are shown in blue, orange, and green respectively. We see that the Cosmic Tunnels are of the order of 1 degree in size, where the RASS and ACT Cosmic Tunnel size distributions peak at $R_v \approx 1\degree$, and the RASS+ACT distribution peaks at $R_v \approx 0.8\degree$. Cosmic Tunnels smaller (larger) than this become increasingly rare, due to the low probability of clusters aligning closely (distantly) in projection.

To first order, the projected number density of clusters in a given catalogue determines the size distribution of the Cosmic Tunnels, and the clustering of the galaxy clusters will play a role through higher-order contributions. The cluster number density is given by the halo mass function, and the galaxy cluster clustering is given by the cluster two-point correlation function. Both of these cluster summary statistics are known to be informative cosmological probes \cite[e.g.][]{Bocquet2024,Fumagalli2024}, and we would therefore also expect that the Cosmic Tunnel size distribution also contains cosmological information. Given that the Cosmic Tunnels are identified over a large redshift range, in most cases the three clusters that define a Cosmic Tunnel are not physically correlated. In this case, we would expect the tunnel size distribution to be entirely determined by the halo mass function. This picture may change in a tomographic analysis, where tunnels are identified from thin redshift slices of the cluster catalogues, hence increasing the cluster-to-cluster correlation, and the cosmological information in the Cosmic Tunnel abundance. However, in this work, we focus on verifying that such objects do indeed correspond to underdense lines of sight to motivate their use in void studies, and we leave a study of the cosmological information to a future work. 

\subsection{Lensing profiles}\label{sec:lensing profiles}
\begin{figure}[!ht]
    \centering
    \begin{subfigure}{\columnwidth}
        \caption{Galaxy WL profiles (DES)}
        \includegraphics[width=\columnwidth]{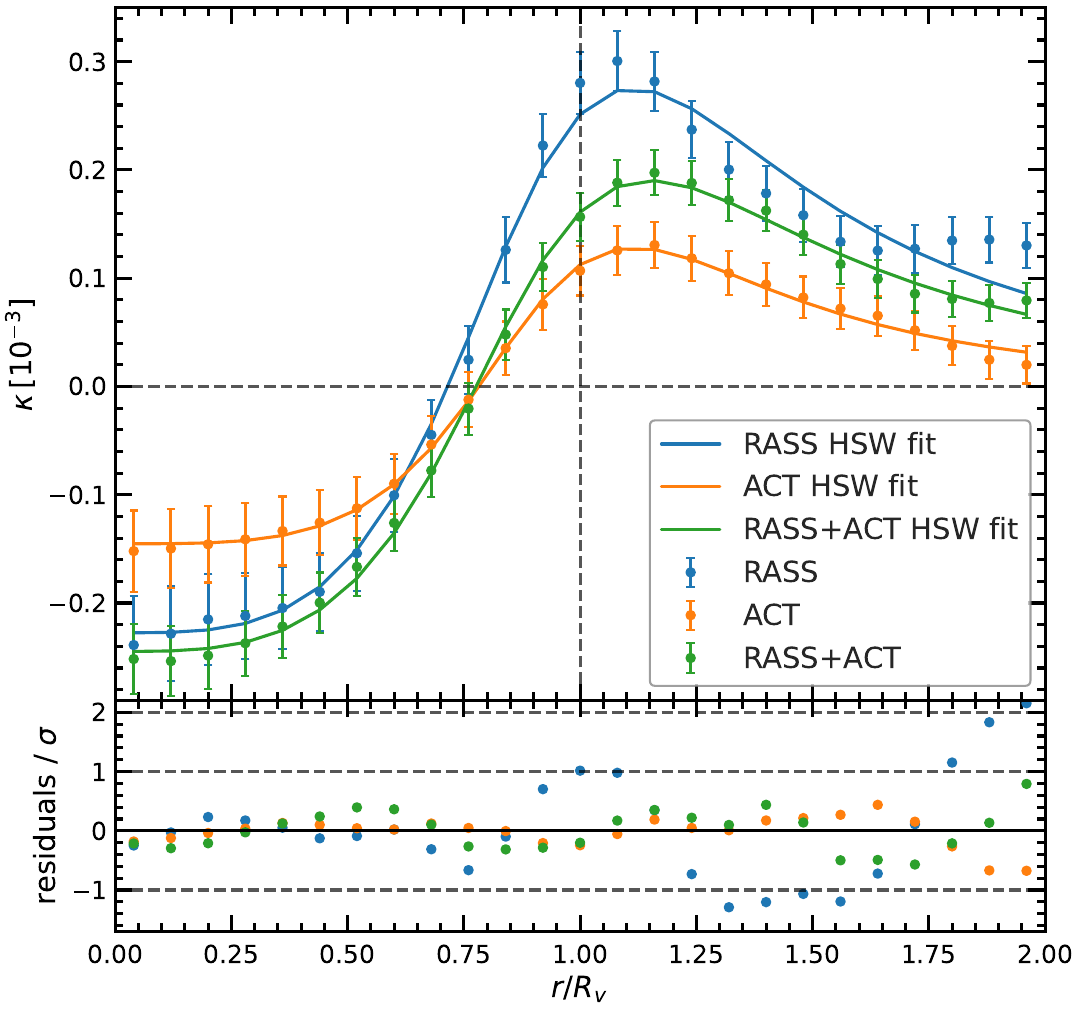}
    \end{subfigure}
    \begin{subfigure}{\columnwidth}
        \caption{CMB WL profiles (ACT)}
        \includegraphics[width=\columnwidth]{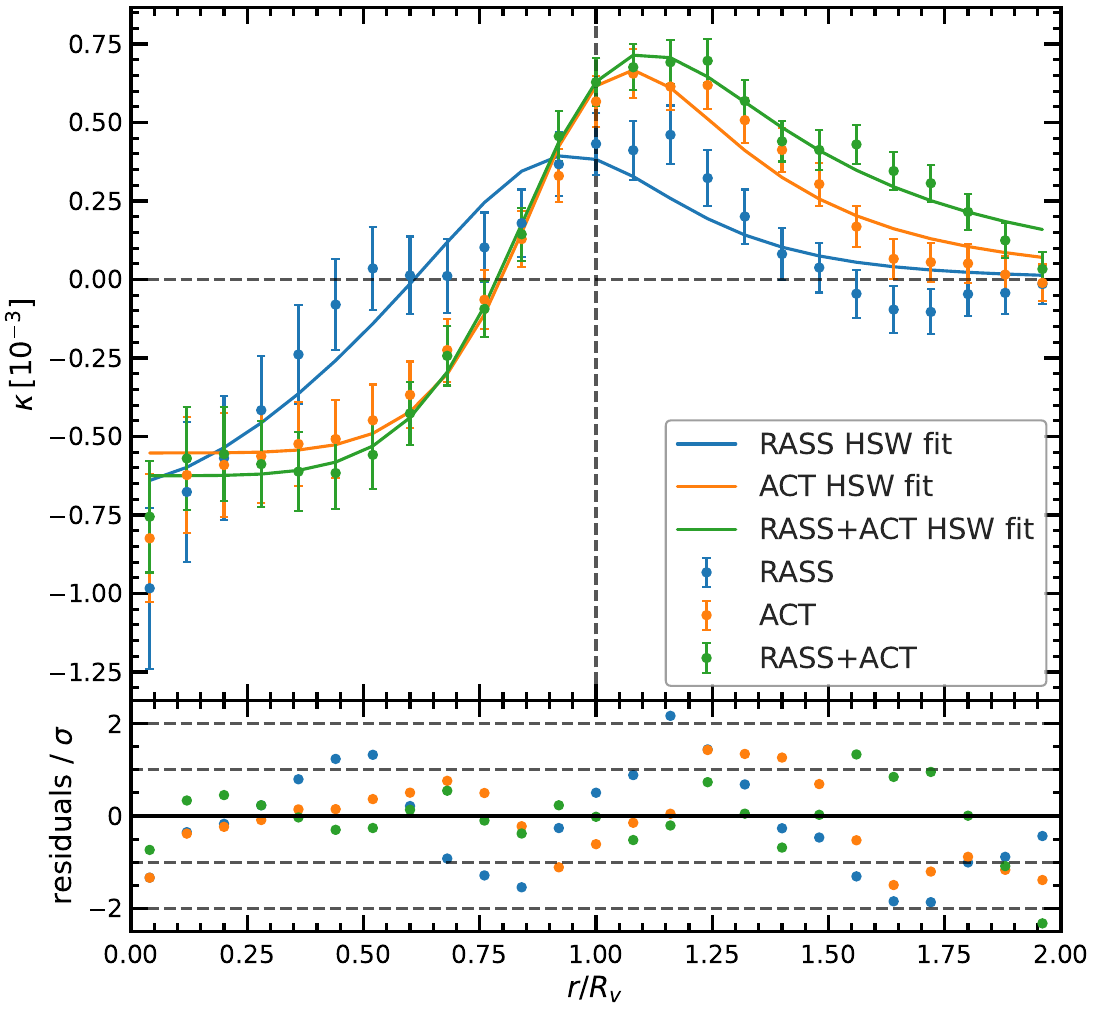}
    \end{subfigure}
    \caption{The radial WL profiles of cosmic tunnels, measured with the DES Y3 mass map (top) and the ACT DR6 mass map (bottom). The WL profiles show the WL convergence $\kappa$ as a function of the distance from the tunnel centre. The data points correspond to the stack of the entire cosmic tunnel population, weighted by void size, for a given cluster catalogue. The blue, orange, and green curves correspond to the Cosmic Tunnels WL profiles measured with the RASS, ACT, and RASS+ACT MCMF cluster catalogues respectively. The error bars indicate the $1\sigma$ standard errors. The solid lines show the HSW fit to the data points, with colours matching the data as shown in the legend. The sub panels show the residuals between the data and the HSW fit. }
    \label{fig:WL profiles}
\end{figure}

In this section, we present two Cosmic Tunnel WL profiles, measured from the DES-Y3 galaxy WL data \citep{Jeffrey2021}, and ACT-DR6 CMB WL data \citep{Madhavacheril2024}, both discussed in Sec. \ref{sec:cross correlation data}. Fig. \ref{fig:WL profiles} shows the galaxy WL profiles (top panel) and CMB WL profiles (bottom panel), for the RASS, ACT, and RASS+ACT Cosmic Tunnels, shown in blue orange and green respectively. The error bars correspond to the $1\sigma$ standard errors estimated from jackknife resampling. We note that the Cosmic Tunnels used here are the same objects as those presented in Fig. \ref{fig:cosmic tunnel abundance}, except objects outside of the corresponding weak-lensing survey footprint do not contribute to the final stacked WL profile.

The figure shows that all WL profiles here are void-like. They all correspond to underdense interiors (negative $\kappa$ values) which gradually increase to overdensities at the boundary (positive $\kappa$ values), and then trend towards the mean density of the Universe ($\kappa$=0) at large distances from the void. 

For galaxy WL, the RASS Cosmic Tunnel WL profile is deeper than the corresponding ACT profile in the interior, and more overdense at the boundary, i.e. the RASS profile has a larger overall amplitude. This is connected to the redshift distribution of the two catalogues, where RASS corresponds to lower redshift, which more closely aligns with the peak of the lensing kernel in the DES-Y3 data \citep{Jeffrey2021}. Therefore, the RASS clusters exist at redshifts where the lesning efficiency is higher in the DES-Y3 data relative to the redshift distribution of the ACT clusters, leading to the larger amplitude for the RASS Cosmic Tunnel WL profile. The same argument can be applied to the CMB WL profiles where the ACT profile is more underdense in the interior and more overdense at the boundary compared to RASS, due to the ACT cluster redshift distribution more closely aligning with the peak of the CMB lensing kernel than RASS. 

For the RASS+ACT Cosmic Tunnel WL profiles, we note that the interior regions are more underdense than both RASS and ACT for both galaxy and CMB WL. This is because the RASS+ACT catalogue contains more tracers, and therefore results in more robust tunnel identification, where spurious tunnel detection would suppress the lensing profiles. For the boundary regions, the RASS+ACT lensing amplitude is larger than RASS and ACT for the CMB WL profile, but for galaxy WL, the RASS+ACT lensing amplitude is larger than ACT, but smaller than RASS. We expect that this is because although more robust tunnels are identified with RASS+ACT, a large number of high-$z$ clusters are added relative to RASS, which lowers the net lensing efficiency of the clusters (which are located at the tunnel boundary), and hence leads to a lower lensing amplitude at the boundary relative to RASS alone. 

Next, we use the method outlined in Appendix. \ref{sec:SNR} to quantify the statistical significance with which the Cosmic Tunnel lensing profiles have been measured. The results are shown in the first two rows of Table \ref{tab:snr}, where we find that the galaxy WL profiles have higher significance than the CMB WL profiles, and the RASS+ACT catalogue yields higher significance than either RASS or ACT alone. We report the strongest detection with the RASS+ACT catalogue galaxy WL profile at $31\sigma$. This verifies that the WL profiles here robustly show that Cosmic Tunnels correspond to underdense lines of sight with very high statistical certainty, and are hence void-like objects.

Next, to further validate the void-like nature of the WL profiles studied here, we fit the HSW profile discussed in Sec. \ref{sec:HSW} to the WL measurements. The results of the HSW fits are shown by the solid lines in Fig. 
\ref{fig:WL profiles}, with colour schemes matching that used for the data. The sub-panels show the corresponding residuals between the data and the fit HSW curves, where the residuals are plotted relative to the error on the data. For the case of galaxy WL, the figure shows that the HSW profile provides a remarkably good fit to all three lensing profiles, where the residuals between the data and the HSW model do not exceed $1\sigma$ for ACT and RASS+ACT over the whole radial range, and remain within $1\sigma$ for RASS below $r/R_v = 1$. At larger radii, the scatter in the residuals for RASS increases to around $2\sigma$ at $r/R_v=2$. We note the presence of oscillatory structure in the residuals, which we attribute to the high bin-to-bin correlation, rather than indicative of a flaw in the model. Nevertheless, this is small for ACT and RASS+ACT, and we find a remarkably strong fit of the HSW model to the galaxy WL data. For CMB WL we find a qualitatively similar fit, although the scatter in the residuals is somewhat larger, but still within $2\sigma$ in nearly all cases. We find this to be consistent with the lower SNRs reported in Table \ref{tab:snr}. Overall, the HSW fits further verify that Cosmic Tunnels are void-like objects.

\subsection{Galaxy profiles}\label{sec:gal profiles}
\begin{figure}
    \centering
    \includegraphics[width=\columnwidth]{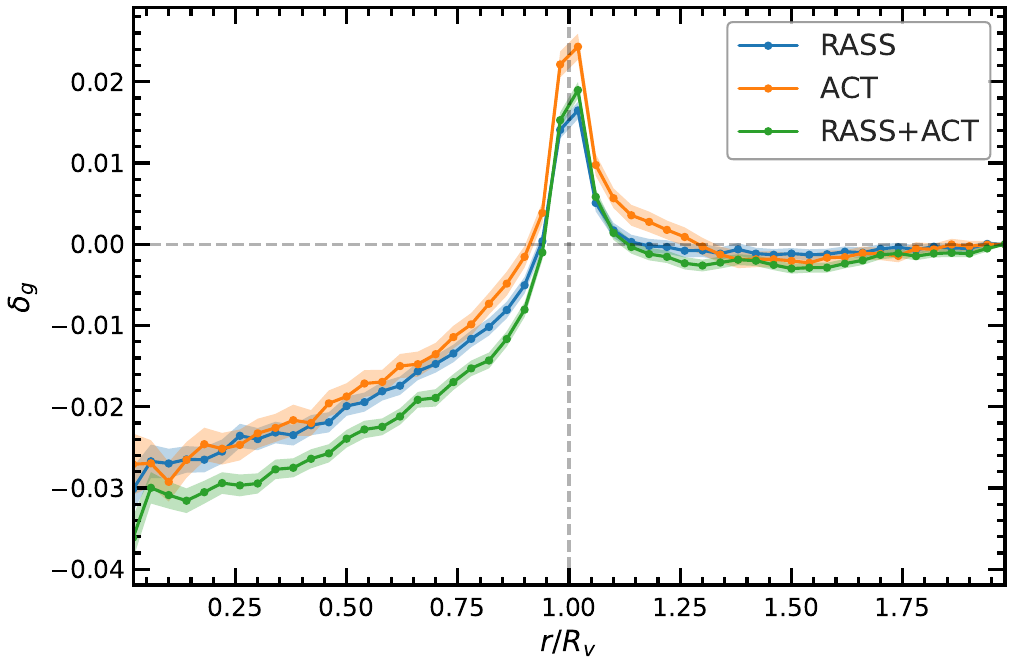}
    \caption{The radial profiles of Cosmic Tunnels in the galaxy field. The profiles are measured in terms of the galaxy overdensity $\delta_g$, and plotted as a function of distance from the tunnel centre normalised by the tunnel radius. The curves shown here correspond to the stack of the entire cosmic tunnel population, weighted by tunnel size, for a given cluster catalogue. Blue, orange, and green show the Cosmic Tunnel galaxy profiles for the RASS, ACT, and RASS+ACT MCMF galaxy cluster catalogues respectively. The shaded regions indicate the $1\sigma$ standard errors. }
    \label{fig:cosmic tunnel galaxy profile}
\end{figure}

In this section, we present the cross correlation between the Cosmic Tunnels and the galaxy projected density contrast field, which can be interpreted as the radial galaxy density profile of the Cosmic Tunnels. The projected galaxy density contrast field used in this study is taken from the Legacy Survey \citep{Legacysurveys19}, which is described in Sec. \ref{sec:cross correlation data}, and we define the galaxy density contrast field according to Eq. \ref{eq:galdencon}. We note that we only use galaxy data in the range $z\in[0,0.5]$, and we therefore apply the same redshift cut to the galaxy cluster catalogues before applying the tunnel algorithm, which results in a different Cosmic Tunnel population to that presented in Fig. \ref{fig:cosmic tunnel abundance}. Due to the inhomogeneous depth across the survey area, we perform a local background subtraction at $r/R_v = 2$ when measuring the cross-correlation with the projected galaxy density field, which fixes the measurements in this region to 0. 

The galaxy density contrast profiles of Cosmic Tunnels identified from the RASS, ACT and RASS+ACT catalogues in the redshift range $z\in[0,0.5]$ are presented in Fig. \ref{fig:cosmic tunnel galaxy profile} as the blue, orange, and green curves respectively. The shaded regions correspond to the 1$\sigma$ standard errors, estimated using the jackknife method. Consistent with the WL lensing profiles, the galaxy density profiles further confirm that the Cosmic Tunnels correspond to underdense lines of sight. For all three catalogues, the profiles exhibit an underdensity in the galaxy field near the tunnel centre, which gradually increases as it approaches the tunnel boundary, which peaks at $r=R_v$, and corresponds to an overdensity in this region. As expected, the profile approaches zero at separations further away from the tunnels, which corresponds to the mean galaxy density contrast. We note that the galaxy density profiles exhibit a sharper peak than the WL profiles at $r=R_v$, due to the lower effective resolution in the WL maps, which is attributed to the fact that the WL measurements are performed on smoothed fields, where no smoothing is applied to the galaxy field after binning galaxy positions onto the \textsc{HEALPix} map.

Consistent with the WL profiles, we find the RASS+ACT catalogue exhibits the greatest underdensity in the galaxy field, which is expected as the RASS+ACT catalogue contains the greatest number of tracers, which gives a more robust probe of the underlying density field, and hence yields fewer spurious tunnels. In contrast to the WL profiles, the ACT catalogue shows the greatest galaxy overdensity. Such differences are expected as the different observables trace the total matter field in different ways, and also have differing trends in redshift, where the ACT catalogue probes higher redshifts than the RASS catalogue. 


\subsection{CMB temperature profiles and the ISW effect}\label{sec:isw}

\begin{figure}
    \centering
    \begin{subfigure}{\columnwidth}
        \includegraphics[width=\columnwidth]{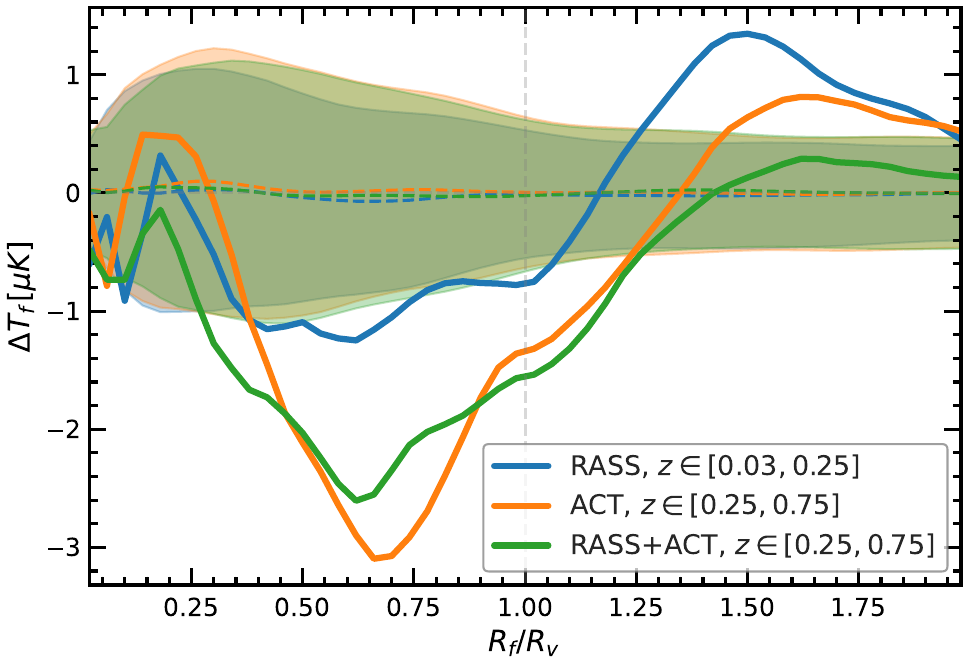}
        \caption{CMB temperature anisotropy profile (CTH filter)}
    \end{subfigure}
    \caption{The CTH radial profiles of Cosmic Tunnels in the CMB temperature anisotropy field. The profiles show the filtered temperature decrement $\Delta T_f$ due to the ISW effect induced by Cosmic Tunnels in the CMB. The profiles are plotted as a function of CTH filter scale ($R_f$) normalised by the tunnel radius. The curves shown here correspond to the mean stack of the Cosmic Tunnels, for the RASS-MCMF (blue) and ACT-MCMF (orange) and RASS+ACT-MCMF (green) cluster catalogues. The dashed lines correspond to the mean value calculated from 1000 mote-carlo realisation from the FFP10 simulations, and the shaded regions indicate the $1\sigma$ standard deviation calculated from the same 1000 simulations. }
    \label{fig:isw}
\end{figure}

In this section, we perform a cross correlation between the Cosmic Tunnels and the CMB temperature anisotropy map from Planck, which is described in Sec. \ref{sec:cross correlation data}. The cross correlation of Cosmic Tunnels with the CMB allows us to probe the ISW effect. The ISW effect is a large-scale secondary anisotropy imprinted on CMB photons as they travel towards the observer from the surface of last scattering. The ISW effect arises from the time evolution of gravitational potentials along the line of sight, below $z\approx1100$, induced by the presence of dark energy \citep{Sachs1967}. 

Due to the low SNR with which the ISW effect can be measured, additional processing of the CMB temperature anisotropy map is typically performed for LSS cross-correlation studies. In this analysis, we employ a compensated top hat filter (CTH) to the CMB map, when measuring the individual Cosmic Tunnel CMB temperature anisotropy profiles. These individual CTH profiles are then stacked, weighted by the Cosmic Tunnel area, to yield the final measurements. The CTH filter is a very common approach employed in ISW studies \cite[e.g.][]{Kovacs2018}. The CTH filter corresponds to the difference between a disk of radius $R_f$ and a ring which encloses that disk of radius $1.4R_f$, centered on the Cosmic Tunnel. The factor of $1.4$ dictates that the disk and the ring have the same area. This means that information from scales larger than $R_f$ are removed from the measurement. The characteristic scale $R_f$ of the CTH filter can then be varied, which in this work we characterise as some factor of the Cosmic Tunnel radius $R_v$. Therefore, for a fixed $R_f$ expressed in units of $R_v$, the physical size of each CTH filter applied to each Cosmic Tunnel before stacking varies, but is held constant relative to the size of the tunnel. This ensures that the final stack correctly aligns the Cosmic Tunnels boundary in the same annuli, which is crucial for robustly detecting the ISW signal. Previous works have shown that the ISW signal from voids is expected to be maximal in the region $R_f \approx 0.7R_v$ \citep{Cai2014}.

Due to the large-scale nature of the ISW effect, the strength of the signal is limited by cosmic variance. For this reason, using the data to quantify the uncertainty in the measurement will lead to an underestimate of the error. To robustly quantify the uncertainty, we use the 1000 random CMB realisations from the FFP10 simulations \cite{Planck2018} to measure the null distribution of the Cosmic Tunnel-CMB temperature cross correlation, where we perform 1000 cross correlations between the true Cosmic Tunnel catalogue and a random CMB realisation. Finally, we quantify the ISW detection significance at the $R_f$ scale where the SNR is maximal. 

The solid lines in Fig. \ref{fig:isw} show the stacked CTH CMB temperature profiles of the Cosmic Tunnels, the dashed lines show the mean null signal measured from the 1000 FFP10 simulations, and the shaded regions show the $1\sigma$ region of the null distribution. The results for RASS are shown in blue, ACT is shown in orange, and RASS+ACT is shown in green. 

For this cross-correlation we select tracer redshift ranges that offer a compromise between probing the epoch where the ISW signal is expected to be present/strongest and the range where the tracer density of the cluster catalogue is highest, which more completely traces the underlying matter field. For these reasons, we use the range $z\in[0.03,0.25]$ for RASS, where the lower limit cuts off late-times (which is discussed in Sec. \ref{sec:lowzisw}), and the upper limit corresponds roughly to the start of the era of dark energy domination. For ACT and RASS+ACT, we use $z\in[0.25,0.75]$, where the lower limit is chosen  not overlap with the RASS measurement, and the upper limit is chosen as the era that roughly corresponds to when dark energy becomes relevant for structure formation. The redshift ranges of the tracers used from each catalogue are also indicated in the figure legend. We expect that the strength of the ISW effect should increase with decreasing redshift, as the contribution to structure formation from dark energy increases. However, in contrast, at low redshift the volume of the Universe is lower, and hence cosmic variance is higher, which reduces the signal-to-noise ratio of the measurement. These two effects act against each other, which we are able to dissect with the above non-overlapping bins. 

Fig. \ref{fig:isw} shows that the shape of the CTH temperature anisotropy profiles matches that expected for standard 3D galaxy voids, with a trough at around $r = 0.7R_v$, as predicted in \cite{Cai2014}. This is consistent with the picture of Cosmic Tunnels as void-like objects established in the previous sections. We therefore attribute the signal measured here as an ISW signal.

The significance with which the ISW signal has been detected is given in Table \ref{tab:snr}, where the maximum significance from RASS, ACT, RASS+ACT is 3.1$\sigma$, 3.6$\sigma$, and 2.7$\sigma$ respectively. These are competitive ISW detection significances measured with a new class of object in the LSS, which opens up a new regime for ISW studies. Note that these significances are taken from individual bins where the SNR is maximal, rather than from the combination of all bins as in the previous sections.

We note that while the ISW detection significance from RASS is high for ISW studies, this measurement corresponds to the signal at the outskirts of the Tunnel, at $R_f \approx 1.5R_v$, which is in contrast to the maximal signal achieved at $R_f \approx 0.7R_v$ with the ACT and RASS+ACT catalogues. We therefore expect that the signal in the RASS measurement is driven by the overdense boundary of the Cosmic Tunnels, which can be attributed to the galaxy clusters. 

We also note that the detection significance from the RASS+ACT catalogue is lower than that in the ACT catalogue, despite the RASS+ACT catalogue containing more tracers, which we would expect to yield a stronger signal. However, when performing the RASS+ACT measurement, we apply both the RASS and ACT mask, thereby reducing the available survey area, which leads to the reduction in amplitude and detection significance. 

We also note that one contaminant to the interpretation of these results as an ISW signal is contributions to the temperature map from the SZ effect. This may be especially prevalent in this analysis, as the SZ signal is strongly associated with galaxy clusters, and especially for SZ-selected clusters. To verify that the signal measured here is indeed due to the ISW effect and not the SZ effect, we repeated the measurement with the ACT Cosmic Tunnels catalogue on the SMICA SZ subtracted temperature anisotropy map. In this case, we find that the signal is further enhanced by 0.2$\sigma$ when SZ is removed, however, we stick to reporting the measurement on the full official Planck data.

\subsection{Low $z$ ISW effect}\label{sec:lowzisw}

\begin{figure}
    \centering
    \includegraphics[width=\columnwidth]{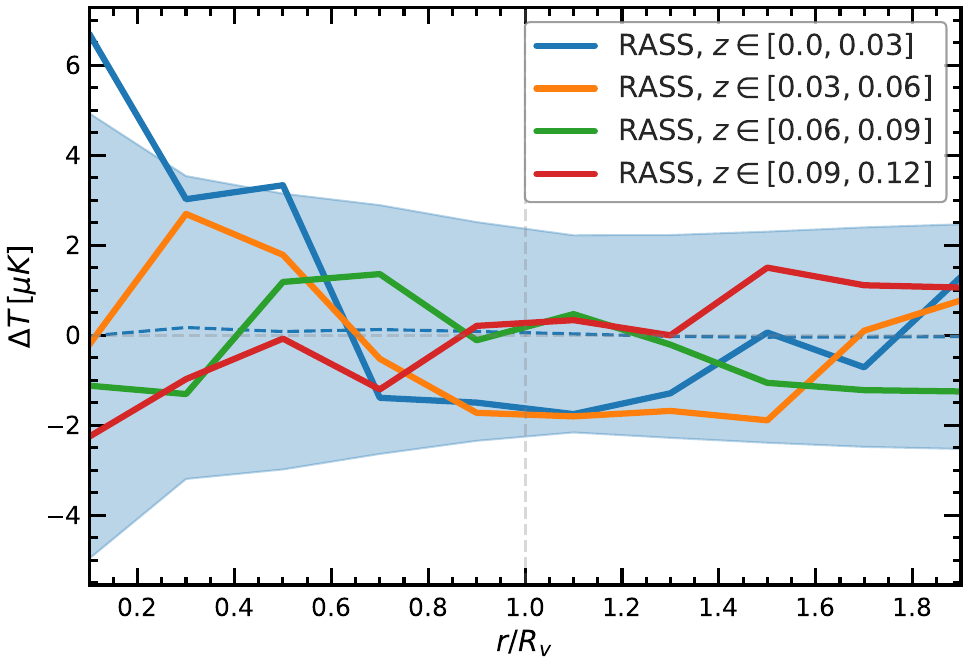}
    \caption{Low redshift CMB temperature anisotropy profiles measured with Cosmic Tunnels from RASS-MCMF. Curves with different colours correspond to different redshifts, as indicated in the legend. The blue dashed line shows the mean measurement from 1000 FFP10 simulations using Cosmic Tunnels from RASS with clusters in the redshift range $z\in[0,0.03]$. The blue shaded region shows the corresponding 1 sigma region on the blue dashed line.}
    \label{fig:lowzisw}
\end{figure}

Finally, we use the methodology outlined in Sec. \ref{sec:isw} to test the measured sign-flip in the ISW effect at $z<0.03$, reported in \cite{Hansen2025}, measured at $3.6\sigma$.
To test this, we identify Cosmic Tunnels using RASS clusters in the redshift range $z\in[0,0.03]$. The mean Cosmic Tunnel radial temperature profile is shown by the blue line in Fig.\ref{fig:lowzisw}, the mean of the null distribution measured from the FFP10 simulations is shown by the dashed line, and the $1\sigma$ width of the null distribution is shown by the shaded region. In this case, we present the unfiltered temperature profile, instead of the CTH profile, as it is simpler to interpret physically, and both the raw temperature profile and CTH profile yield the same significance in this regime. The figure shows that we also detect a sign flip in the ISW effect at very low redshift, where the expectation in $\Lambda$CDM would be to find a temperature decrement, but we instead find a temperature enhancement. We note however, that the statistical significance with which we measure this sign flip is only $1.4\sigma$, which is significantly lower than that reported in \cite{Hansen2025}. One factor behind the lower significance reported here is that additional size cuts are applied to the low redshift voids in \cite{Hansen2025}, which leads to approximately a $1\sigma$ increase in detection significance, whereas in this work, we do not apply any additional size cuts. At $1.4\sigma$ we are unable to rule out a chance fluctuation as the cause of the ISW sign flip. 

Given the higher $z$ results in the previous section, where the Cosmic Tunnel ISW signal corresponds to a decrement in the CMB temperature as expected, we investigate additional tomographic slices. This allows us to test whether the Cosmic Tunnel CMB profile transitions from a temperature decrement to a temperature enhancement as a function of $z$, which can provide further justification for the sign-flip narrative. We find that in the adjacent redshift bin, $z\in[0.03,0.06]$, the sign flip is still present, and the amplitude has decreased. The decrease in amplitude reduces the statistical significance of the sign flip in this bin, but remains consistent with the narrative that the ISW effect transitions from a temperature decrement to an enhancement with decreasing $z$. The higher redshift bins are also consistent with the null signal, well within $1\sigma$ \footnote{We note that for each tomographic slice, a corresponding null distribution should be calculated with the FFP10 simulations. However, as the null distribution does not change significantly in Fig. \ref{fig:isw}, we expect the error bar presented for the first tomographic bin in Fig. \ref{fig:lowzisw} to be broadly representative of all tomographic bins presented here.}, although they consistently trend towards a temperature decrement at higher $z$. It is possible that this in itself is a signature of new ISW physics, as in the standard $\Lambda$CDM paradigm, the ISW signal should be stronger at lower redshift. Although at this stage we cannot rule out that the low detection significance is due to the lower volumes at low redshift, and hence greater sample variance. We therefore caution that all of these results are statistically consistent with the null signal.

\section{Conclusions}

In this work we have defined a new void-like object, Cosmic Tunnels, which correspond to large underdense lines of sight traced by galaxy clusters. To identify these objects in data we apply the Tunnel algorithm from \cite{Cautun2018} to two galaxy cluster catalogues, RASS-MCMF \citep{RASSMCMF} and ACT-MCMF \citep{klein2024actdr5}. 

We test the underdense and void-like nature of Cosmic Tunnels through cross-correlations with galaxy weak lensing from DES Y3 \citep{Jeffrey2021}, CMB weak lensing from ACT DR6 \citep{Madhavacheril2024}, and the galaxy distribution from the DESI Legacy Survey \cite{Legacysurveys19}. Through these cross-correlations we verify that Cosmic Tunnels are underdense, and futhermore, we find that the cosmic tunnel weak lensing profiles can be fit with the universal HSW profile, providing further motivation to treat these objects as void-like. We measure the Cosmic Tunnel galaxy weak lensing profiles at $30\sigma$ significance, the CMB weak lensing profiles at $15\sigma$ significance, and the galaxy profiles at $60\sigma$ significance. 

To exploit the Cosmic Tunnels, we perform a cross-correlation with the Planck temperature anisotropy map to detect the ISW signal. We apply a CTH filter to the Planck map when measuring the profiles, and investigate the redshift range $z\in[0.25,0.75]$ (similar to other ISW studies \citep[][e.g.]{Nadathur2012}). We find that the CMB temperature-Cosmic Tunnel cross-correlation does indeed yield a signal consistent with the ISW effect, with a peak at CTH filter scale $R_f = 0.7R_v$ matching expectations from previous studies (e.g. \citep{Cai2014}). We use the Planck FFP10 simulations to quantify the statistical significance with which the ISW signal was detected, finding a $3.6\sigma$ detection significance with the ACT Cosmic Tunnel catalogue. We also perform a low redshift ($z$<0.03) study of the ISW effect, and find tentative evidence of a sign flip in the ISW effect, which would be inconsistent with $\Lambda$CDM, although we caution that this is measured with low statistical significance. 

The high statistical significance of the cross-correlations performed in this work opens a new regime for using under-densities to probe the Universe. A joint analysis of the weak lensing and galaxy cross-correlations can be used to constrain the galaxy-halo connection, and individual and joint ISW and weak lensing analyses can be used to test $\Lambda\rm{CDM}$ in new regimes. 

Given that Cosmic Tunnel abundances are directly tied to the projected cluster distribution, their size function may also offer a new route for cosmological parameter inference, similar to the cosmological information contained within the 3D void size function \citep{Contarini2023}. A forward model of the Cosmic Tunnel abundance based on the cluster halo mass functions could therefore allow future analyses to extract cosmological parameter constraints from tunnel statistics alone.

Cosmic Tunnels complement existing 2D and 3D void catalogues by probing a distinct tracer population with minimal redshift uncertainty. Combining Cosmic Tunnel based voids with weak lensing \citep{Davies2020b} or galaxy \citep{Contarini2023} defined voids could enable joint cosmological analyses that cross-validate and constrain systematic uncertainties, such as the galaxy-halo connection and redshift uncertainties.

Finally, further methods for identifying Cosmic Tunnels may include applications to optical cluster catalogues, which further sample complementary cluster populations as with the SZ and X-ray selected cluster catalogues used in this work. Such an analysis with optical cluster catalogues may further benefit from the increased number density of optical clusters relative to SZ and X-ray selected clusters.

\begin{acknowledgements}
We acknowledge support from the Ludwig Maximilians-Universität in Munich
\end{acknowledgements}

%
%
\bibliographystyle{aa}
\bibliography{bib}

\begin{thebibliography}{70}
\expandafter\ifx\csname natexlab\endcsname\relax\def\natexlab#1{#1}\fi

\bibitem[{{Abbott} {et~al.}(2016){Abbott}, {Abdalla}, {Aleksi{\'c}}, {Allam}, {Amara}, {Bacon}, {Balbinot}, {Banerji}, {Bechtol}, {Benoit-L{\'e}vy}, {Bernstein}, {Bertin}, {Blazek}, {Bonnett}, {Bridle}, {Brooks}, {Brunner}, {Buckley-Geer}, {Burke}, {Caminha}, {Capozzi}, {Carlsen}, {Carnero-Rosell}, {Carollo}, {Carrasco-Kind}, {Carretero}, {Castander}, {Clerkin}, {Collett}, {Conselice}, {Crocce}, {Cunha}, {D'Andrea}, {da Costa}, {Davis}, {Desai}, {Diehl}, {Dietrich}, {Dodelson}, {Doel}, {Drlica-Wagner}, {Estrada}, {Etherington}, {Evrard}, {Fabbri}, {Finley}, {Flaugher}, {Foley}, {Fosalba}, {Frieman}, {Garc{\'\i}a-Bellido}, {Gaztanaga}, {Gerdes}, {Giannantonio}, {Goldstein}, {Gruen}, {Gruendl}, {Guarnieri}, {Gutierrez}, {Hartley}, {Honscheid}, {Jain}, {James}, {Jeltema}, {Jouvel}, {Kessler}, {King}, {Kirk}, {Kron}, {Kuehn}, {Kuropatkin}, {Lahav}, {Li}, {Lima}, {Lin}, {Maia}, {Makler}, {Manera}, {Maraston}, {Marshall}, {Martini}, {McMahon}, {Melchior}, {Merson}, {Miller}, {Miquel}, {Mohr}, {Morice-Atkinson},
  {Naidoo}, {Neilsen}, {Nichol}, {Nord}, {Ogando}, {Ostrovski}, {Palmese}, {Papadopoulos}, {Peiris}, {Peoples}, {Percival}, {Plazas}, {Reed}, {Refregier}, {Romer}, {Roodman}, {Ross}, {Rozo}, {Rykoff}, {Sadeh}, {Sako}, {S{\'a}nchez}, {Sanchez}, {Santiago}, {Scarpine}, {Schubnell}, {Sevilla-Noarbe}, {Sheldon}, {Smith}, {Smith}, {Soares-Santos}, {Sobreira}, {Soumagnac}, {Suchyta}, {Sullivan}, {Swanson}, {Tarle}, {Thaler}, {Thomas}, {Thomas}, {Tucker}, {Vieira}, {Vikram}, {Walker}, {Wechsler}, {Weller}, {Wester}, {Whiteway}, {Wilcox}, {Yanny}, {Zhang}, \& {Zuntz}}]{DES2016}
{Abbott}, T., {Abdalla}, F.~B., {Aleksi{\'c}}, J., {et~al.} 2016, \mnras, 460, 1270

\bibitem[{Achitouv(2016)}]{Achitouv2016}
Achitouv, I. 2016, Phys. Rev. D, 94, 103524

\bibitem[{Anderson(2003)}]{Anderson2003}
Anderson, T.~W. 2003, An introduction to multivariate statistical analysis (Wiley-Interscience)

\bibitem[{Barreira {et~al.}(2015)Barreira, Cautun, Li, Baugh, \& Pascoli}]{Barreira2015}
Barreira, A., Cautun, M., Li, B., Baugh, C.~M., \& Pascoli, S. 2015, \jcap, 2015, 028

\bibitem[{{Bleem} {et~al.}(2024){Bleem}, {Klein}, {Abbot}, {Ade}, {Aguena}, {Alves}, {Anderson}, {Andrade-Oliveira}, {Ansarinejad}, {Archipley}, {Ashby}, {Austermann}, {Bacon}, {Beall}, {Bender}, {Benson}, {Bianchini}, {Bocquet}, {Brooks}, {Burke}, {Calzadilla}, {Carlstrom}, {Carnero Rosell}, {Carretero}, {Chang}, {Chaubal}, {Chiang}, {Chou}, {Citron}, {Corbett Moran}, {Costanzi}, {Constanzi}, {Crawford}, {Crites}, {da Costa}, {de Haan}, {De Vicente}, {Desai}, {Dobbs}, {Doel}, {Everett}, {Ferrero}, {Flaugher}, {Floyd}, {Friedel}, {Frieman}, {Gallicchio}, {Garc'ia-Bellido}, {Gatti}, {George}, {Giannini}, {Grandis}, {Gruen}, {Gruendl}, {Gupta}, {Gutierrez}, {Halverson}, {Hinton}, {Hinton}, {Holder}, {Hollowood}, {Holzapfel}, {Honscheid}, {Hrubes}, {Huang}, {Hubmayr}, {Irwin}, {Mena-Fern{\'a}ndez}, {James}, {K{\'e}ruzor{\'e}}, {Knox}, {Kuehn}, {Lahav}, {Lee}, {Lee}, {Li}, {Lowitz}, {Marshal}, {McDonald}, {McMahon}, {Menanteau}, {Meyer}, {Miquel}, {Mohr}, {Montgomery}, {Myles}, {Natoli}, {Nibarger}, {Noble},
  {Novosad}, {Ogando}, {Padin}, {Patil}, {Pereira}, {Pieres}, {Plazas Malag'on}, {Pryke}, {Reichardt}, {Rodr'iguez-Monroy}, {Romer}, {Ruhl}, {Saliwanchik}, {Salvati}, {Sanchez}, {Saro}, {Schaffer}, {Schrabback}, {Sevilla-Noarbe}, {Sievers}, {Smecher}, {Smith}, {Somboonpanyakul}, {Stalder}, {Stark}, {Suchyta}, {Swanson}, {Tarle}, {To}, {Tucker}, {Veach}, {Vieira}, {Vincenzi}, {Wang}, {Weller}, {Whitehorn}, {Wiseman}, {Wu}, {Yefremenko}, {Zebrowski}, \& {Zhang}}]{2024OJAp....7E..13B}
{Bleem}, L.~E., {Klein}, M., {Abbot}, T.~M.~C., {et~al.} 2024, The Open Journal of Astrophysics, 7, 13

\bibitem[{{Bocquet} {et~al.}(2024){Bocquet}, {Grandis}, {Bleem}, {Klein}, {Mohr}, {Schrabback}, {Abbott}, {Ade}, {Aguena}, {Alarcon}, {Allam}, {Allen}, {Alves}, {Amon}, {Anderson}, {Annis}, {Ansarinejad}, {Austermann}, {Avila}, {Bacon}, {Bayliss}, {Beall}, {Bechtol}, {Becker}, {Bender}, {Benson}, {Bernstein}, {Bhargava}, {Bianchini}, {Brodwin}, {Brooks}, {Bryant}, {Campos}, {Canning}, {Carlstrom}, {Carnero Rosell}, {Carrasco Kind}, {Carretero}, {Castander}, {Cawthon}, {Chang}, {Chang}, {Chaubal}, {Chen}, {Chiang}, {Choi}, {Chou}, {Citron}, {Corbett Moran}, {Cordero}, {Costanzi}, {Crawford}, {Crites}, {da Costa}, {Pereira}, {Davis}, {Davis}, {DeRose}, {Desai}, {de Haan}, {Diehl}, {Dobbs}, {Dodelson}, {Doux}, {Drlica-Wagner}, {Eckert}, {Elvin-Poole}, {Everett}, {Everett}, {Ferrero}, {Fert{\'e}}, {Flores}, {Frieman}, {Gallicchio}, {Garc{\'\i}a-Bellido}, {Gatti}, {George}, {Giannini}, {Gladders}, {Gruen}, {Gruendl}, {Gupta}, {Gutierrez}, {Halverson}, {Harrison}, {Hartley}, {Herner}, {Hinton}, {Holder},
  {Hollowood}, {Holzapfel}, {Honscheid}, {Hrubes}, {Huang}, {Hubmayr}, {Huff}, {Huterer}, {Irwin}, {James}, {Jarvis}, {Khullar}, {Kim}, {Knox}, {Kraft}, {Krause}, {Kuehn}, {Kuropatkin}, {K{\'e}ruzor{\'e}}, {Lahav}, {Lee}, {Leget}, {Li}, {Lin}, {Lowitz}, {MacCrann}, {Mahler}, {Mantz}, {Marshall}, {McCullough}, {McDonald}, {McMahon}, {Mena-Fern{\'a}ndez}, {Menanteau}, {Meyer}, {Miquel}, {Montgomery}, {Myles}, {Natoli}, {Navarro-Alsina}, {Nibarger}, {Noble}, {Novosad}, {Ogando}, {Omori}, {Padin}, {Pandey}, {Paschos}, {Patil}, {Pieres}, {Plazas Malag{\'o}n}, {Porredon}, {Prat}, {Pryke}, {Raveri}, {Reichardt}, {Roberson}, {Rollins}, {Romero}, {Roodman}, {Ruhl}, {Rykoff}, {Saliwanchik}, {Salvati}, {S{\'a}nchez}, {Sanchez}, {Sanchez Cid}, {Saro}, {Schaffer}, {Secco}, {Sevilla-Noarbe}, {Sharon}, {Sheldon}, {Shin}, {Sievers}, {Smecher}, {Smith}, {Somboonpanyakul}, {Sommer}, {Stalder}, {Stark}, {Stephen}, {Strazzullo}, {Suchyta}, {Tarle}, {To}, {Troxel}, {Tucker}, {Tutusaus}, {Varga}, {Veach}, {Vieira}, {Vikhlinin},
  {von der Linden}, {Wang}, {Weaverdyck}, {Weller}, {Whitehorn}, {Wu}, {Yanny}, {Yefremenko}, {Yin}, {Young}, {Zebrowski}, {Zhang}, {Zohren}, {Zuntz}, {(SPT}, \& {DES Collaborations)}}]{Bocquet2024}
{Bocquet}, S., {Grandis}, S., {Bleem}, L.~E., {et~al.} 2024, \prd, 110, 083510

\bibitem[{{Boller} {et~al.}(2016){Boller}, {Freyberg}, {Tr{\"u}mper}, {Haberl}, {Voges}, \& {Nandra}}]{Boller2RXS}
{Boller}, T., {Freyberg}, M.~J., {Tr{\"u}mper}, J., {et~al.} 2016, \aap, 588, A103

\bibitem[{{Bond} {et~al.}(1996){Bond}, {Kofman}, \& {Pogosyan}}]{Bond1996}
{Bond}, J.~R., {Kofman}, L., \& {Pogosyan}, D. 1996, \nat, 380, 603

\bibitem[{{Cai} {et~al.}(2014){Cai}, {Neyrinck}, {Szapudi}, {Cole}, \& {Frenk}}]{Cai2014}
{Cai}, Y.-C., {Neyrinck}, M.~C., {Szapudi}, I., {Cole}, S., \& {Frenk}, C.~S. 2014, \apj, 786, 110

\bibitem[{{Cautun} {et~al.}(2018){Cautun}, {Paillas}, {Cai}, {Bose}, {Armijo}, {Li}, \& {Padilla}}]{Cautun2018}
{Cautun}, M., {Paillas}, E., {Cai}, Y.-C., {et~al.} 2018, Mon. Not. Roy. Astron. Soc., 476, 3195

\bibitem[{{Cautun} {et~al.}(2013){Cautun}, {van de Weygaert}, \& {Jones}}]{Cautun2013}
{Cautun}, M., {van de Weygaert}, R., \& {Jones}, B. J.~T. 2013, Mon. Not. Roy. Astron. Soc., 429, 1286

\bibitem[{{Colberg} {et~al.}(2008){Colberg}, {Pearce}, {Foster}, {Platen}, {Brunino}, {Neyrinck}, {Basilakos}, {Fairall}, {Feldman}, {Gottl{\"o}ber}, {Hahn}, {Hoyle}, {M{\"u}ller}, {Nelson}, {Plionis}, {Porciani}, {Shandarin}, {Vogeley}, \& {van de Weygaert}}]{Colberg2008}
{Colberg}, J.~M., {Pearce}, F., {Foster}, C., {et~al.} 2008, Mon. Not. Roy. Astron. Soc., 387, 933

\bibitem[{{Contarini} {et~al.}(2023){Contarini}, {Pisani}, {Hamaus}, {Marulli}, {Moscardini}, \& {Baldi}}]{Contarini2023}
{Contarini}, S., {Pisani}, A., {Hamaus}, N., {et~al.} 2023, \apj, 953, 46

\bibitem[{{Contarini} {et~al.}(2019){Contarini}, {Ronconi}, {Marulli}, {Moscardini}, {Veropalumbo}, \& {Baldi}}]{Contarini2019}
{Contarini}, S., {Ronconi}, T., {Marulli}, F., {et~al.} 2019, \mnras, 488, 3526

\bibitem[{{Davies} {et~al.}(2021){Davies}, {Cautun}, {Giblin}, {Li}, {Harnois-D{\'e}raps}, \& {Cai}}]{Davies2020b}
{Davies}, C.~T., {Cautun}, M., {Giblin}, B., {et~al.} 2021, \mnras, 507, 2267

\bibitem[{{Davies} {et~al.}(2018){Davies}, {Cautun}, \& {Li}}]{Davies2018}
{Davies}, C.~T., {Cautun}, M., \& {Li}, B. 2018, \mnras, 480, L101

\bibitem[{{Davies} {et~al.}(2019){Davies}, {Cautun}, \& {Li}}]{Davies2019b}
{Davies}, C.~T., {Cautun}, M., \& {Li}, B. 2019, \mnras, 490, 4907

\bibitem[{{Davies} {et~al.}(2020){Davies}, {Paillas}, {Cautun}, \& {Li}}]{Davies2020}
{Davies}, C.~T., {Paillas}, E., {Cautun}, M., \& {Li}, B. 2020, arXiv e-prints, arXiv:2004.11387

\bibitem[{{Davis} {et~al.}(1985){Davis}, {Efstathiou}, {Frenk}, \& {White}}]{Davis1985}
{Davis}, M., {Efstathiou}, G., {Frenk}, C.~S., \& {White}, S.~D.~M. 1985, Astrophys. J., 292, 371

\bibitem[{Del~Giudice(2021)}]{DelGiudice2021}
Del~Giudice, M. 2021, Multivariate Behavioral Research, 56, 527

\bibitem[{{Demirbozan} {et~al.}(2024){Demirbozan}, {Nadathur}, {Ferrero}, {Fosalba}, {Kov{\'a}cs}, {Miquel}, {Davies}, {Pandey}, {Adamow}, {Bechtol}, {Drlica-Wagner}, {Gruendl}, {Hartley}, {Pieres}, {Ross}, {Rykoff}, {Sheldon}, {Yanny}, {Abbott}, {Aguena}, {Allam}, {Alves}, {Bacon}, {Bertin}, {Bocquet}, {Brooks}, {Rosell}, {Carretero}, {Cawthon}, {da Costa}, {Pereira}, {De Vicente}, {Desai}, {Doel}, {Everett}, {Flaugher}, {Friedel}, {Frieman}, {Gatti}, {Gaztanaga}, {Giannini}, {Gutierrez}, {Hinton}, {Hollowood}, {James}, {Jeffrey}, {Kuehn}, {Lahav}, {Lee}, {Marshall}, {Mena-Fern{\'a}ndez}, {Mohr}, {Myles}, {Ogando}, {Malag{\'o}n}, {Roodman}, {Sanchez}, {Sevilla-Noarbe}, {Smith}, {Soares-Santos}, {Suchyta}, {Swanson}, {Tarle}, {Weaverdyck}, {Weller}, \& {Wiseman}}]{Demirbozan2024}
{Demirbozan}, U., {Nadathur}, S., {Ferrero}, I., {et~al.} 2024, \mnras, 534, 2328

\bibitem[{{Dey} {et~al.}(2019){Dey}, {Schlegel}, {Lang}, {Blum}, {Burleigh}, {Fan}, {Findlay}, {Finkbeiner}, {Herrera}, {Juneau}, {Landriau}, {Levi}, {McGreer}, {Meisner}, {Myers}, {Moustakas}, {Nugent}, {Patej}, {Schlafly}, {Walker}, {Valdes}, {Weaver}, {Y{\`e}che}, {Zou}, {Zhou}, {Abareshi}, {Abbott}, {Abolfathi}, {Aguilera}, {Alam}, {Allen}, {Alvarez}, {Annis}, {Ansarinejad}, {Aubert}, {Beechert}, {Bell}, {BenZvi}, {Beutler}, {Bielby}, {Bolton}, {Brice{\~n}o}, {Buckley-Geer}, {Butler}, {Calamida}, {Carlberg}, {Carter}, {Casas}, {Castander}, {Choi}, {Comparat}, {Cukanovaite}, {Delubac}, {DeVries}, {Dey}, {Dhungana}, {Dickinson}, {Ding}, {Donaldson}, {Duan}, {Duckworth}, {Eftekharzadeh}, {Eisenstein}, {Etourneau}, {Fagrelius}, {Farihi}, {Fitzpatrick}, {Font-Ribera}, {Fulmer}, {G{\"a}nsicke}, {Gaztanaga}, {George}, {Gerdes}, {Gontcho}, {Gorgoni}, {Green}, {Guy}, {Harmer}, {Hernandez}, {Honscheid}, {Huang}, {James}, {Jannuzi}, {Jiang}, {Joyce}, {Karcher}, {Karkar}, {Kehoe}, {Kneib}, {Kueter-Young}, {Lan},
  {Lauer}, {Le Guillou}, {Le Van Suu}, {Lee}, {Lesser}, {Perreault Levasseur}, {Li}, {Mann}, {Marshall}, {Mart{\'\i}nez-V{\'a}zquez}, {Martini}, {du Mas des Bourboux}, {McManus}, {Meier}, {M{\'e}nard}, {Metcalfe}, {Mu{\~n}oz-Guti{\'e}rrez}, {Najita}, {Napier}, {Narayan}, {Newman}, {Nie}, {Nord}, {Norman}, {Olsen}, {Paat}, {Palanque-Delabrouille}, {Peng}, {Poppett}, {Poremba}, {Prakash}, {Rabinowitz}, {Raichoor}, {Rezaie}, {Robertson}, {Roe}, {Ross}, {Ross}, {Rudnick}, {Safonova}, {Saha}, {S{\'a}nchez}, {Savary}, {Schweiker}, {Scott}, {Seo}, {Shan}, {Silva}, {Slepian}, {Soto}, {Sprayberry}, {Staten}, {Stillman}, {Stupak}, {Summers}, {Sien Tie}, {Tirado}, {Vargas-Maga{\~n}a}, {Vivas}, {Wechsler}, {Williams}, {Yang}, {Yang}, {Yapici}, {Zaritsky}, {Zenteno}, {Zhang}, {Zhang}, {Zhou}, \& {Zhou}}]{Legacysurveys19}
{Dey}, A., {Schlegel}, D.~J., {Lang}, D., {et~al.} 2019, \aj, 157, 168

\bibitem[{{Drlica-Wagner} {et~al.}(2022){Drlica-Wagner}, {Ferguson}, {Adam{\'o}w}, {Aguena}, {Allam}, {Andrade-Oliveira}, {Bacon}, {Bechtol}, {Bell}, {Bertin}, {Bilaji}, {Bocquet}, {Bom}, {Brooks}, {Burke}, {Carballo-Bello}, {Carlin}, {Carnero Rosell}, {Carrasco Kind}, {Carretero}, {Castander}, {Cerny}, {Chang}, {Choi}, {Conselice}, {Costanzi}, {Crnojevi{\'c}}, {da Costa}, {de Vicente}, {Desai}, {Esteves}, {Everett}, {Ferrero}, {Fitzpatrick}, {Flaugher}, {Friedel}, {Frieman}, {Garc{\'\i}a-Bellido}, {Gatti}, {Gaztanaga}, {Gerdes}, {Gruen}, {Gruendl}, {Gschwend}, {Hartley}, {Hernandez-Lang}, {Hinton}, {Hollowood}, {Honscheid}, {Hughes}, {Jacques}, {James}, {Johnson}, {Kuehn}, {Kuropatkin}, {Lahav}, {Li}, {Lidman}, {Lin}, {March}, {Marshall}, {Mart{\'\i}nez-Delgado}, {Mart{\'\i}nez-V{\'a}zquez}, {Massana}, {Mau}, {McNanna}, {Melchior}, {Menanteau}, {Miller}, {Miquel}, {Mohr}, {Morgan}, {Mutlu-Pakdil}, {Mu{\~n}oz}, {Neilsen}, {Nidever}, {Nikutta}, {Nilo Castellon}, {No{\"e}l}, {Ogando}, {Olsen}, {Pace},
  {Palmese}, {Paz-Chinch{\'o}n}, {Pereira}, {Pieres}, {Plazas Malag{\'o}n}, {Prat}, {Riley}, {Rodriguez-Monroy}, {Romer}, {Roodman}, {Sako}, {Sakowska}, {Sanchez}, {S{\'a}nchez}, {Sand}, {Santana-Silva}, {Santiago}, {Schubnell}, {Serrano}, {Sevilla-Noarbe}, {Simon}, {Smith}, {Soares-Santos}, {Stringfellow}, {Suchyta}, {Suson}, {Tan}, {Tarle}, {Tavangar}, {Thomas}, {To}, {Tollerud}, {Troxel}, {Tucker}, {Varga}, {Vivas}, {Walker}, {Weller}, {Wilkinson}, {Wu}, {Yanny}, {Zaborowski}, {Zenteno}, {Delve Collaboration}, {Des Collaboration}, \& {Astro Data Lab}}]{DELVE}
{Drlica-Wagner}, A., {Ferguson}, P.~S., {Adam{\'o}w}, M., {et~al.} 2022, \apjs, 261, 38

\bibitem[{Fang {et~al.}(2019)Fang, Hamaus, Jain, Pandey, Pollina, Sánchez, Kovács, Chang, Carretero, Castander, Choi, Crocce, DeRose, Fosalba, Gatti, Gaztañaga, Gruen, Hartley, Hoyle, MacCrann, Prat, Rau, Rykoff, Samuroff, Sheldon, Troxel, Vielzeuf, Zuntz, Annis, Avila, Bertin, Brooks, Burke, Carnero Rosell, Carrasco Kind, Cawthon, da Costa, De Vicente, Desai, Diehl, Dietrich, Doel, Everett, Evrard, Flaugher, Frieman, García-Bellido, Gerdes, Gruendl, Gutierrez, Hollowood, James, Jarvis, Kuropatkin, Lahav, Maia, Marshall, Melchior, Menanteau, Miquel, Palmese, Plazas, Romer, Roodman, Sanchez, Serrano, Sevilla-Noarbe, Smith, Soares-Santos, Sobreira, Suchyta, Swanson, Tarle, Thomas, Vikram, Walker, Weller, \& Collaboration)}]{Fang2019}
Fang, Y., Hamaus, N., Jain, B., {et~al.} 2019, Mon. Not. Roy. Astron. Soc., 490, 3573

\bibitem[{{Flaugher} {et~al.}(2015){Flaugher}, {Diehl}, {Honscheid}, {Abbott}, {Alvarez}, {Angstadt}, {Annis}, {Antonik}, {Ballester}, {Beaufore}, {Bernstein}, {Bernstein}, {Bigelow}, {Bonati}, {Boprie}, {Brooks}, {Buckley-Geer}, {Campa}, {Cardiel-Sas}, {Castander}, {Castilla}, {Cease}, {Cela-Ruiz}, {Chappa}, {Chi}, {Cooper}, {da Costa}, {Dede}, {Derylo}, {DePoy}, {de Vicente}, {Doel}, {Drlica-Wagner}, {Eiting}, {Elliott}, {Emes}, {Estrada}, {Fausti Neto}, {Finley}, {Flores}, {Frieman}, {Gerdes}, {Gladders}, {Gregory}, {Gutierrez}, {Hao}, {Holland}, {Holm}, {Huffman}, {Jackson}, {James}, {Jonas}, {Karcher}, {Karliner}, {Kent}, {Kessler}, {Kozlovsky}, {Kron}, {Kubik}, {Kuehn}, {Kuhlmann}, {Kuk}, {Lahav}, {Lathrop}, {Lee}, {Levi}, {Lewis}, {Li}, {Mandrichenko}, {Marshall}, {Martinez}, {Merritt}, {Miquel}, {Mu{\~n}oz}, {Neilsen}, {Nichol}, {Nord}, {Ogando}, {Olsen}, {Palaio}, {Patton}, {Peoples}, {Plazas}, {Rauch}, {Reil}, {Rheault}, {Roe}, {Rogers}, {Roodman}, {Sanchez}, {Scarpine}, {Schindler}, {Schmidt},
  {Schmitt}, {Schubnell}, {Schultz}, {Schurter}, {Scott}, {Serrano}, {Shaw}, {Smith}, {Soares-Santos}, {Stefanik}, {Stuermer}, {Suchyta}, {Sypniewski}, {Tarle}, {Thaler}, {Tighe}, {Tran}, {Tucker}, {Walker}, {Wang}, {Watson}, {Weaverdyck}, {Wester}, {Woods}, {Yanny}, \& {DES Collaboration}}]{Flaugher15}
{Flaugher}, B., {Diehl}, H.~T., {Honscheid}, K., {et~al.} 2015, \aj, 150, 150

\bibitem[{{Fowler} {et~al.}(2007){Fowler}, {Niemack}, {Dicker}, {Aboobaker}, {Ade}, {Battistelli}, {Devlin}, {Fisher}, {Halpern}, {Hargrave}, {Hincks}, {Kaul}, {Klein}, {Lau}, {Limon}, {Marriage}, {Mauskopf}, {Page}, {Staggs}, {Swetz}, {Switzer}, {Thornton}, \& {Tucker}}]{ACTtel}
{Fowler}, J.~W., {Niemack}, M.~D., {Dicker}, S.~R., {et~al.} 2007, \ao, 46, 3444

\bibitem[{{Fumagalli} {et~al.}(2024){Fumagalli}, {Costanzi}, {Saro}, {Castro}, \& {Borgani}}]{Fumagalli2024}
{Fumagalli}, A., {Costanzi}, M., {Saro}, A., {Castro}, T., \& {Borgani}, S. 2024, \aap, 682, A148

\bibitem[{{Granett} {et~al.}(2008){Granett}, {Neyrinck}, \& {Szapudi}}]{Granett2008}
{Granett}, B.~R., {Neyrinck}, M.~C., \& {Szapudi}, I. 2008, \apjl, 683, L99

\bibitem[{Gruen {et~al.}(2015)Gruen, Friedrich, Amara, Bacon, Bonnett, Hartley, Jain, Jarvis, Kacprzak, Krause, Mana, Rozo, Rykoff, Seitz, Sheldon, Troxel, Vikram, Abbott, Abdalla, Allam, Armstrong, Banerji, Bauer, Becker, Benoit-Lévy, Bernstein, Bernstein, Bertin, Bridle, Brooks, Buckley-Geer, Burke, Capozzi, Carnero~Rosell, Carrasco~Kind, Carretero, Crocce, Cunha, D'Andrea, da~Costa, DePoy, Desai, Diehl, Dietrich, Doel, Eifler, Neto, Fernandez, Flaugher, Fosalba, Frieman, Gerdes, Gruendl, Gutierrez, Honscheid, James, Kuehn, Kuropatkin, Lahav, Li, Lima, Maia, March, Martini, Melchior, Miller, Miquel, Mohr, Nord, Ogando, Plazas, Reil, Romer, Roodman, Sako, Sanchez, Scarpine, Schubnell, Sevilla-Noarbe, Smith, Soares-Santos, Sobreira, Suchyta, Swanson, Tarle, Thaler, Thomas, Walker, Wechsler, Weller, Zhang, \& Zuntz}]{Gruen2015}
Gruen, D., Friedrich, O., Amara, A., {et~al.} 2015, Mon. Not. Roy. Astron. Soc., 455, 3367

\bibitem[{{Hamaus} {et~al.}(2016){Hamaus}, {Pisani}, {Sutter}, {Lavaux}, {Escoffier}, {Wand elt}, \& {Weller}}]{Hamaus2016}
{Hamaus}, N., {Pisani}, A., {Sutter}, P.~M., {et~al.} 2016, Phys. Rev. Lett., 117, 091302

\bibitem[{{Hamaus} {et~al.}(2014){Hamaus}, {Sutter}, \& {Wandelt}}]{Hamaus2014}
{Hamaus}, N., {Sutter}, P.~M., \& {Wandelt}, B.~D. 2014, \prl, 112, 251302

\bibitem[{{Hang} {et~al.}(2021){Hang}, {Alam}, {Cai}, \& {Peacock}}]{Hang2021}
{Hang}, Q., {Alam}, S., {Cai}, Y.-C., \& {Peacock}, J.~A. 2021, \mnras, 507, 510

\bibitem[{{Hansen} {et~al.}(2025){Hansen}, {Garcia Lambas}, {Ruiz}, {Toscano}, \& {Pereyra}}]{Hansen2025}
{Hansen}, F.~K., {Garcia Lambas}, D., {Ruiz}, A.~N., {Toscano}, F., \& {Pereyra}, L.~A. 2025, arXiv e-prints, arXiv:2506.08832

\bibitem[{{Hilton} {et~al.}(2021){Hilton}, {Sif{\'o}n}, {Naess}, {Madhavacheril}, {Oguri}, {Rozo}, {Rykoff}, {Abbott}, {Adhikari}, {Aguena}, {Aiola}, {Allam}, {Amodeo}, {Amon}, {Annis}, {Ansarinejad}, {Aros-Bunster}, {Austermann}, {Avila}, {Bacon}, {Battaglia}, {Beall}, {Becker}, {Bernstein}, {Bertin}, {Bhandarkar}, {Bhargava}, {Bond}, {Brooks}, {Burke}, {Calabrese}, {Carrasco Kind}, {Carretero}, {Choi}, {Choi}, {Conselice}, {da Costa}, {Costanzi}, {Crichton}, {Crowley}, {D{\"u}nner}, {Denison}, {Devlin}, {Dicker}, {Diehl}, {Dietrich}, {Doel}, {Duff}, {Duivenvoorden}, {Dunkley}, {Everett}, {Ferraro}, {Ferrero}, {Fert{\'e}}, {Flaugher}, {Frieman}, {Gallardo}, {Garc{\'\i}a-Bellido}, {Gaztanaga}, {Gerdes}, {Giles}, {Golec}, {Gralla}, {Grandis}, {Gruen}, {Gruendl}, {Gschwend}, {Gutierrez}, {Han}, {Hartley}, {Hasselfield}, {Hill}, {Hilton}, {Hincks}, {Hinton}, {Ho}, {Honscheid}, {Hoyle}, {Hubmayr}, {Huffenberger}, {Hughes}, {Jaelani}, {Jain}, {James}, {Jeltema}, {Kent}, {Knowles}, {Koopman}, {Kuehn}, {Lahav},
  {Lima}, {Lin}, {Lokken}, {Loubser}, {MacCrann}, {Maia}, {Marriage}, {Martin}, {McMahon}, {Melchior}, {Menanteau}, {Miquel}, {Miyatake}, {Moodley}, {Morgan}, {Mroczkowski}, {Nati}, {Newburgh}, {Niemack}, {Nishizawa}, {Ogando}, {Orlowski-Scherer}, {Page}, {Palmese}, {Partridge}, {Paz-Chinch{\'o}n}, {Phakathi}, {Plazas}, {Robertson}, {Romer}, {Carnero Rosell}, {Salatino}, {Sanchez}, {Schaan}, {Schillaci}, {Sehgal}, {Serrano}, {Shin}, {Simon}, {Smith}, {Soares-Santos}, {Spergel}, {Staggs}, {Storer}, {Suchyta}, {Swanson}, {Tarle}, {Thomas}, {To}, {Trac}, {Ullom}, {Vale}, {Van Lanen}, {Vavagiakis}, {De Vicente}, {Wilkinson}, {Wollack}, {Xu}, \& {Zhang}}]{ACTDR5}
{Hilton}, M., {Sif{\'o}n}, C., {Naess}, S., {et~al.} 2021, \apjs, 253, 3

\bibitem[{{Jeffrey} {et~al.}(2021){Jeffrey}, {Gatti}, {Chang}, {Whiteway}, {Demirbozan}, {Kovacs}, {Pollina}, {Bacon}, {Hamaus}, {Kacprzak}, {Lahav}, {Lanusse}, {Mawdsley}, {Nadathur}, {Starck}, {Vielzeuf}, {Zeurcher}, {Alarcon}, {Amon}, {Bechtol}, {Bernstein}, {Campos}, {Rosell}, {Kind}, {Cawthon}, {Chen}, {Choi}, {Cordero}, {Davis}, {DeRose}, {Doux}, {Drlica-Wagner}, {Eckert}, {Elsner}, {Elvin-Poole}, {Everett}, {Fert{\'e}}, {Giannini}, {Gruen}, {Gruendl}, {Harrison}, {Hartley}, {Herner}, {Huff}, {Huterer}, {Kuropatkin}, {Jarvis}, {Leget}, {MacCrann}, {McCullough}, {Muir}, {Myles}, {Navarro-Alsina}, {Pandey}, {Prat}, {Raveri}, {Rollins}, {Ross}, {Rykoff}, {S{\'a}nchez}, {Secco}, {Sevilla-Noarbe}, {Sheldon}, {Shin}, {Troxel}, {Tutusaus}, {Varga}, {Yanny}, {Yin}, {Zhang}, {Zuntz}, {Abbott}, {Aguena}, {Allam}, {Andrade-Oliveira}, {Becker}, {Bertin}, {Bhargava}, {Brooks}, {Burke}, {Carretero}, {Castander}, {Conselice}, {Costanzi}, {Crocce}, {da Costa}, {Pereira}, {De Vicente}, {Desai}, {Diehl}, {Dietrich},
  {Doel}, {Ferrero}, {Flaugher}, {Fosalba}, {Garc{\'\i}a-Bellido}, {Gaztanaga}, {Gerdes}, {Giannantonio}, {Gschwend}, {Gutierrez}, {Hinton}, {Hollowood}, {Hoyle}, {Jain}, {James}, {Lima}, {Maia}, {March}, {Marshall}, {Melchior}, {Menanteau}, {Miquel}, {Mohr}, {Morgan}, {Ogando}, {Palmese}, {Paz-Chinch{\'o}n}, {Plazas}, {Rodriguez-Monroy}, {Roodman}, {Sanchez}, {Scarpine}, {Serrano}, {Smith}, {Soares-Santos}, {Suchyta}, {Tarle}, {Thomas}, {To}, {Weller}, \& {DES Collaboration}}]{Jeffrey2021}
{Jeffrey}, N., {Gatti}, M., {Chang}, C., {et~al.} 2021, \mnras, 505, 4626

\bibitem[{{Kilbinger}(2015)}]{Kilbinger2015}
{Kilbinger}, M. 2015, Rept. Prog. Phys., 78, 086901

\bibitem[{{Kirshner} {et~al.}(1981){Kirshner}, {Oemler}, {Schechter}, \& {Shectman}}]{Kirshner1981}
{Kirshner}, R.~P., {Oemler}, Jr., A., {Schechter}, P.~L., \& {Shectman}, S.~A. 1981, Astrophys. J.l, 248, L57

\bibitem[{{Klein} {et~al.}(2019){Klein}, {Grandis}, {Mohr}, {Paulus}, {Abbott}, {Annis}, {Avila}, {Bertin}, {Brooks}, {Buckley-Geer}, {Rosell}, {Kind}, {Carretero}, {Castander}, {Cunha}, {D'Andrea}, {da Costa}, {De Vicente}, {Desai}, {Diehl}, {Dietrich}, {Doel}, {Evrard}, {Flaugher}, {Fosalba}, {Frieman}, {Garc{\'\i}a-Bellido}, {Gaztanaga}, {Giles}, {Gruen}, {Gruendl}, {Gschwend}, {Gutierrez}, {Hartley}, {Hollowood}, {Honscheid}, {Hoyle}, {James}, {Jeltema}, {Kuehn}, {Kuropatkin}, {Lima}, {Maia}, {March}, {Marshall}, {Menanteau}, {Miquel}, {Ogando}, {Plazas}, {Romer}, {Roodman}, {Sanchez}, {Scarpine}, {Schindler}, {Serrano}, {Sevilla-Noarbe}, {Smith}, {Smith}, {Soares-Santos}, {Sobreira}, {Suchyta}, {Swanson}, {Tarle}, {Thomas}, {Vikram}, \& {DES Collaboration}}]{MARDY3}
{Klein}, M., {Grandis}, S., {Mohr}, J.~J., {et~al.} 2019, \mnras, 488, 739

\bibitem[{{Klein} {et~al.}(2023){Klein}, {Hern{\'a}ndez-Lang}, {Mohr}, {Bocquet}, \& {Singh}}]{RASSMCMF}
{Klein}, M., {Hern{\'a}ndez-Lang}, D., {Mohr}, J.~J., {Bocquet}, S., \& {Singh}, A. 2023, \mnras, 526, 3757

\bibitem[{{Klein} {et~al.}(2024{\natexlab{a}}){Klein}, {Mohr}, {Bocquet}, {Aguena}, {Allen}, {Alves}, {Ansarinejad}, {Ashby}, {Bacon}, {Bayliss}, {Benson}, {Bleem}, {Brodwin}, {Brooks}, {Bulbul}, {Burke}, {Canning}, {Carlstrom}, {Rosell}, {Carretero}, {Chang}, {Conselice}, {Costanzi}, {Crites}, {da Costa}, {Pereira}, {Davis}, {De Vicente}, {Desai}, {de Haan}, {Dobbs}, {Doel}, {Ferrero}, {Flores}, {Frieman}, {George}, {Giannini}, {Gladders}, {Gonzalez}, {Grandis}, {Gruen}, {Gruendl}, {Gutierrez}, {Halverson}, {Hinton}, {Holder}, {Hollowood}, {Holzapfel}, {Honscheid}, {Hrubes}, {Huang}, {James}, {Khullar}, {Kim}, {Knox}, {Kraft}, {K{\'e}ruzor{\'e}}, {Lee}, {Luong-Van}, {Mahler}, {Mantz}, {Marrone}, {Marshall}, {McDonald}, {McMahon}, {Mena-Fern{\'a}ndez}, {Menanteau}, {Meyer}, {Miquel}, {Myles}, {Padin}, {Pieres}, {Plazas Malag{\'o}n}, {Pryke}, {Reichardt}, {Reil}, {Roberson}, {Romer}, {Romero}, {Ruhl}, {Saliwanchik}, {Salvati}, {Sanchez}, {Saro}, {Schaffer}, {Schrabback}, {Schubnell}, {Sevilla-Noarbe},
  {Sharon}, {Shirokoff}, {Smith}, {Somboonpanyakul}, {Stalder}, {Stanford}, {Stark}, {Strazzullo}, {Suchyta}, {Swanson}, {Tarle}, {To}, {Vanderlinde}, {Vieira}, {von der Linden}, {Weaverdyck}, {Williamson}, {Wiseman}, \& {Young}}]{SPTSZMCMF}
{Klein}, M., {Mohr}, J.~J., {Bocquet}, S., {et~al.} 2024{\natexlab{a}}, \mnras, 531, 3973

\bibitem[{{Klein} {et~al.}(2024{\natexlab{b}}){Klein}, {Mohr}, \& {Davies}}]{klein2024actdr5}
{Klein}, M., {Mohr}, J.~J., \& {Davies}, C.~T. 2024{\natexlab{b}}, \aap, 690, A322

\bibitem[{{Klein} {et~al.}(2018){Klein}, {Mohr}, {Desai}, {Israel}, {Allam}, {Benoit-L{\'e}vy}, {Brooks}, {Buckley-Geer}, {Carnero Rosell}, {Carrasco Kind}, {Cunha}, {da Costa}, {Dietrich}, {Eifler}, {Evrard}, {Frieman}, {Gruen}, {Gruendl}, {Gutierrez}, {Honscheid}, {James}, {Kuehn}, {Lima}, {Maia}, {March}, {Melchior}, {Menanteau}, {Miquel}, {Plazas}, {Reil}, {Romer}, {Sanchez}, {Santiago}, {Scarpine}, {Schubnell}, {Sevilla-Noarbe}, {Smith}, {Soares-Santos}, {Sobreira}, {Suchyta}, {Swanson}, {Tarle}, \& {DES Collaboration}}]{2018MNRAS.474.3324K}
{Klein}, M., {Mohr}, J.~J., {Desai}, S., {et~al.} 2018, \mnras, 474, 3324

\bibitem[{{Kov{\'a}cs}(2018)}]{Kovacs2018}
{Kov{\'a}cs}, A. 2018, \mnras, 475, 1777

\bibitem[{{Kreisch} {et~al.}(2019){Kreisch}, {Pisani}, {Carbone}, {Liu}, {Hawken}, {Massara}, {Spergel}, \& {Wandelt}}]{Kreisch2019}
{Kreisch}, C.~D., {Pisani}, A., {Carbone}, C., {et~al.} 2019, \mnras, 488, 4413

\bibitem[{{Krolewski} {et~al.}(2018){Krolewski}, {Lee}, {White}, {Hennawi}, {Schlegel}, {Nugent}, {Luki{\'c}}, {Stark}, {Koekemoer}, {Le F{\`e}vre}, {Lemaux}, {Maier}, {Rich}, {Salvato}, \& {Tasca}}]{Krolewski2018}
{Krolewski}, A., {Lee}, K.-G., {White}, M., {et~al.} 2018, \apj, 861, 60

\bibitem[{{Lavaux} \& {Wandelt}(2012)}]{Lavaux2012}
{Lavaux}, G. \& {Wandelt}, B.~D. 2012, Astrophys. J., 754, 109

\bibitem[{{Madhavacheril} {et~al.}(2024){Madhavacheril}, {Qu}, {Sherwin}, {MacCrann}, {Li}, {Abril-Cabezas}, {Ade}, {Aiola}, {Alford}, {Amiri}, {Amodeo}, {An}, {Atkins}, {Austermann}, {Battaglia}, {Battistelli}, {Beall}, {Bean}, {Beringue}, {Bhandarkar}, {Biermann}, {Bolliet}, {Bond}, {Cai}, {Calabrese}, {Calafut}, {Capalbo}, {Carrero}, {Challinor}, {Chesmore}, {Cho}, {Choi}, {Clark}, {C{\'o}rdova Rosado}, {Cothard}, {Coughlin}, {Coulton}, {Crowley}, {Dalal}, {Darwish}, {Devlin}, {Dicker}, {Doze}, {Duell}, {Duff}, {Duivenvoorden}, {Dunkley}, {D{\"u}nner}, {Fanfani}, {Fankhanel}, {Farren}, {Ferraro}, {Freundt}, {Fuzia}, {Gallardo}, {Garrido}, {Givans}, {Gluscevic}, {Golec}, {Guan}, {Hall}, {Halpern}, {Han}, {Harrison}, {Hasselfield}, {Healy}, {Henderson}, {Hensley}, {Herv{\'\i}as-Caimapo}, {Hill}, {Hilton}, {Hilton}, {Hincks}, {Hlo{\v{z}}ek}, {Ho}, {Huber}, {Hubmayr}, {Huffenberger}, {Hughes}, {Irwin}, {Isopi}, {Jense}, {Keller}, {Kim}, {Knowles}, {Koopman}, {Kosowsky}, {Kramer}, {Kusiak}, {La Posta}, {Lague},
  {Lakey}, {Lee}, {Li}, {Limon}, {Lokken}, {Louis}, {Lungu}, {MacInnis}, {Maldonado}, {Maldonado}, {Mallaby-Kay}, {Marques}, {McMahon}, {Mehta}, {Menanteau}, {Moodley}, {Morris}, {Mroczkowski}, {Naess}, {Namikawa}, {Nati}, {Newburgh}, {Nicola}, {Niemack}, {Nolta}, {Orlowski-Scherer}, {Page}, {Pandey}, {Partridge}, {Prince}, {Puddu}, {Radiconi}, {Robertson}, {Rojas}, {Sakuma}, {Salatino}, {Schaan}, {Schmitt}, {Sehgal}, {Shaikh}, {Sierra}, {Sievers}, {Sif{\'o}n}, {Simon}, {Sonka}, {Spergel}, {Staggs}, {Storer}, {Switzer}, {Tampier}, {Thornton}, {Trac}, {Treu}, {Tucker}, {Ullom}, {Vale}, {Van Engelen}, {Van Lanen}, {van Marrewijk}, {Vargas}, {Vavagiakis}, {Wagoner}, {Wang}, {Wenzl}, {Wollack}, {Xu}, {Zago}, \& {Zheng}}]{Madhavacheril2024}
{Madhavacheril}, M.~S., {Qu}, F.~J., {Sherwin}, B.~D., {et~al.} 2024, \apj, 962, 113

\bibitem[{{Maggiore} {et~al.}(2025){Maggiore}, {Contarini}, {Giocoli}, \& {Moscardini}}]{Maggiore2025}
{Maggiore}, L., {Contarini}, S., {Giocoli}, C., \& {Moscardini}, L. 2025, \aap, 701, A55

\bibitem[{{Mao} {et~al.}(2017){Mao}, {Berlind}, {Scherrer}, {Neyrinck}, {Scoccimarro}, {Tinker}, {McBride}, {Schneider}, {Pan}, {Bizyaev}, {Malanushenko}, \& {Malanushenko}}]{Mao2017}
{Mao}, Q., {Berlind}, A.~A., {Scherrer}, R.~J., {et~al.} 2017, Astrophys. J., 835, 161

\bibitem[{Massara {et~al.}(2015)Massara, Villaescusa-Navarro, Viel, \& Sutter}]{Massara2015}
Massara, E., Villaescusa-Navarro, F., Viel, M., \& Sutter, P. 2015, \jcap, 2015, 018

\bibitem[{{Nadathur} {et~al.}(2012){Nadathur}, {Hotchkiss}, \& {Sarkar}}]{Nadathur2012}
{Nadathur}, S., {Hotchkiss}, S., \& {Sarkar}, S. 2012, \jcap, 2012, 042

\bibitem[{{Naess} {et~al.}(2020){Naess}, {Aiola}, {Austermann}, {Battaglia}, {Beall}, {Becker}, {Bond}, {Calabrese}, {Choi}, {Cothard}, {Crowley}, {Darwish}, {Datta}, {Denison}, {Devlin}, {Duell}, {Duff}, {Duivenvoorden}, {Dunkley}, {D{\"u}nner}, {Fox}, {Gallardo}, {Halpern}, {Han}, {Hasselfield}, {Hill}, {Hilton}, {Hilton}, {Hincks}, {Hlo{\v{z}}ek}, {Ho}, {Hubmayr}, {Huffenberger}, {Hughes}, {Kosowsky}, {Louis}, {Madhavacheril}, {McMahon}, {Moodley}, {Nati}, {Nibarger}, {Niemack}, {Page}, {Partridge}, {Salatino}, {Schaan}, {Schillaci}, {Schmitt}, {Sherwin}, {Sehgal}, {Sif{\'o}n}, {Spergel}, {Staggs}, {Stevens}, {Storer}, {Ullom}, {Vale}, {Van Engelen}, {Van Lanen}, {Vavagiakis}, {Wollack}, \& {Xu}}]{ACTDR5data}
{Naess}, S., {Aiola}, S., {Austermann}, J.~E., {et~al.} 2020, \jcap, 12, 046

\bibitem[{Padilla {et~al.}(2005)Padilla, Ceccarelli, \& Lambas}]{Padilla2005}
Padilla, N.~D., Ceccarelli, L., \& Lambas, D.~G. 2005, Mon. Not. Roy. Astron. Soc., 363, 977

\bibitem[{{Paillas} {et~al.}(2019){Paillas}, {Cautun}, {Li}, {Cai}, {Padilla}, {Armijo}, \& {Bose}}]{Paillas2019}
{Paillas}, E., {Cautun}, M., {Li}, B., {et~al.} 2019, Mon. Not. Roy. Astron. Soc., 484, 1149

\bibitem[{Pisani {et~al.}(2015)Pisani, Sutter, Hamaus, Alizadeh, Biswas, Wandelt, \& Hirata}]{Pisani2015}
Pisani, A., Sutter, P.~M., Hamaus, N., {et~al.} 2015, Phys. Rev. D, 92, 083531

\bibitem[{{Planck Collaboration} {et~al.}(2020{\natexlab{a}}){Planck Collaboration}, {Aghanim}, {Akrami}, {Ashdown}, {Aumont}, {Baccigalupi}, {Ballardini}, {Banday}, {Barreiro}, {Bartolo}, {Basak}, {Battye}, {Benabed}, {Bernard}, {Bersanelli}, {Bielewicz}, {Bock}, {Bond}, {Borrill}, {Bouchet}, {Boulanger}, {Bucher}, {Burigana}, {Butler}, {Calabrese}, {Cardoso}, {Carron}, {Challinor}, {Chiang}, {Chluba}, {Colombo}, {Combet}, {Contreras}, {Crill}, {Cuttaia}, {de Bernardis}, {de Zotti}, {Delabrouille}, {Delouis}, {Di Valentino}, {Diego}, {Dor{\'e}}, {Douspis}, {Ducout}, {Dupac}, {Dusini}, {Efstathiou}, {Elsner}, {En{\ss}lin}, {Eriksen}, {Fantaye}, {Farhang}, {Fergusson}, {Fernandez-Cobos}, {Finelli}, {Forastieri}, {Frailis}, {Fraisse}, {Franceschi}, {Frolov}, {Galeotta}, {Galli}, {Ganga}, {G{\'e}nova-Santos}, {Gerbino}, {Ghosh}, {Gonz{\'a}lez-Nuevo}, {G{\'o}rski}, {Gratton}, {Gruppuso}, {Gudmundsson}, {Hamann}, {Handley}, {Hansen}, {Herranz}, {Hildebrandt}, {Hivon}, {Huang}, {Jaffe}, {Jones}, {Karakci},
  {Keih{\"a}nen}, {Keskitalo}, {Kiiveri}, {Kim}, {Kisner}, {Knox}, {Krachmalnicoff}, {Kunz}, {Kurki-Suonio}, {Lagache}, {Lamarre}, {Lasenby}, {Lattanzi}, {Lawrence}, {Le Jeune}, {Lemos}, {Lesgourgues}, {Levrier}, {Lewis}, {Liguori}, {Lilje}, {Lilley}, {Lindholm}, {L{\'o}pez-Caniego}, {Lubin}, {Ma}, {Mac{\'\i}as-P{\'e}rez}, {Maggio}, {Maino}, {Mandolesi}, {Mangilli}, {Marcos-Caballero}, {Maris}, {Martin}, {Martinelli}, {Mart{\'\i}nez-Gonz{\'a}lez}, {Matarrese}, {Mauri}, {McEwen}, {Meinhold}, {Melchiorri}, {Mennella}, {Migliaccio}, {Millea}, {Mitra}, {Miville-Desch{\^e}nes}, {Molinari}, {Montier}, {Morgante}, {Moss}, {Natoli}, {N{\o}rgaard-Nielsen}, {Pagano}, {Paoletti}, {Partridge}, {Patanchon}, {Peiris}, {Perrotta}, {Pettorino}, {Piacentini}, {Polastri}, {Polenta}, {Puget}, {Rachen}, {Reinecke}, {Remazeilles}, {Renzi}, {Rocha}, {Rosset}, {Roudier}, {Rubi{\~n}o-Mart{\'\i}n}, {Ruiz-Granados}, {Salvati}, {Sandri}, {Savelainen}, {Scott}, {Shellard}, {Sirignano}, {Sirri}, {Spencer}, {Sunyaev}, {Suur-Uski},
  {Tauber}, {Tavagnacco}, {Tenti}, {Toffolatti}, {Tomasi}, {Trombetti}, {Valenziano}, {Valiviita}, {Van Tent}, {Vibert}, {Vielva}, {Villa}, {Vittorio}, {Wandelt}, {Wehus}, {White}, {White}, {Zacchei}, \& {Zonca}}]{Planck2018}
{Planck Collaboration}, {Aghanim}, N., {Akrami}, Y., {et~al.} 2020{\natexlab{a}}, \aap, 641, A6

\bibitem[{{Planck Collaboration} {et~al.}(2020{\natexlab{b}}){Planck Collaboration}, {Aghanim}, {Akrami}, {Ashdown}, {Aumont}, {Baccigalupi}, {Ballardini}, {Banday}, {Barreiro}, {Bartolo}, {Basak}, {Benabed}, {Bernard}, {Bersanelli}, {Bielewicz}, {Bock}, {Bond}, {Borrill}, {Bouchet}, {Boulanger}, {Bucher}, {Burigana}, {Butler}, {Calabrese}, {Cardoso}, {Carron}, {Casaponsa}, {Challinor}, {Chiang}, {Colombo}, {Combet}, {Crill}, {Cuttaia}, {de Bernardis}, {de Rosa}, {de Zotti}, {Delabrouille}, {Delouis}, {Di Valentino}, {Diego}, {Dor{\'e}}, {Douspis}, {Ducout}, {Dupac}, {Dusini}, {Efstathiou}, {Elsner}, {En{\ss}lin}, {Eriksen}, {Fantaye}, {Fernandez-Cobos}, {Finelli}, {Frailis}, {Fraisse}, {Franceschi}, {Frolov}, {Galeotta}, {Galli}, {Ganga}, {G{\'e}nova-Santos}, {Gerbino}, {Ghosh}, {Giraud-H{\'e}raud}, {Gonz{\'a}lez-Nuevo}, {G{\'o}rski}, {Gratton}, {Gruppuso}, {Gudmundsson}, {Hamann}, {Handley}, {Hansen}, {Herranz}, {Hivon}, {Huang}, {Jaffe}, {Jones}, {Keih{\"a}nen}, {Keskitalo}, {Kiiveri}, {Kim}, {Kisner},
  {Krachmalnicoff}, {Kunz}, {Kurki-Suonio}, {Lagache}, {Lamarre}, {Lasenby}, {Lattanzi}, {Lawrence}, {Le Jeune}, {Levrier}, {Lewis}, {Liguori}, {Lilje}, {Lilley}, {Lindholm}, {L{\'o}pez-Caniego}, {Lubin}, {Ma}, {Mac{\'\i}as-P{\'e}rez}, {Maggio}, {Maino}, {Mandolesi}, {Mangilli}, {Marcos-Caballero}, {Maris}, {Martin}, {Mart{\'\i}nez-Gonz{\'a}lez}, {Matarrese}, {Mauri}, {McEwen}, {Meinhold}, {Melchiorri}, {Mennella}, {Migliaccio}, {Millea}, {Miville-Desch{\^e}nes}, {Molinari}, {Moneti}, {Montier}, {Morgante}, {Moss}, {Natoli}, {N{\o}rgaard-Nielsen}, {Pagano}, {Paoletti}, {Partridge}, {Patanchon}, {Peiris}, {Perrotta}, {Pettorino}, {Piacentini}, {Polenta}, {Puget}, {Rachen}, {Reinecke}, {Remazeilles}, {Renzi}, {Rocha}, {Rosset}, {Roudier}, {Rubi{\~n}o-Mart{\'\i}n}, {Ruiz-Granados}, {Salvati}, {Sandri}, {Savelainen}, {Scott}, {Shellard}, {Sirignano}, {Sirri}, {Spencer}, {Sunyaev}, {Suur-Uski}, {Tauber}, {Tavagnacco}, {Tenti}, {Toffolatti}, {Tomasi}, {Trombetti}, {Valiviita}, {Van Tent}, {Vielva}, {Villa},
  {Vittorio}, {Wandelt}, {Wehus}, {Zacchei}, \& {Zonca}}]{Planck2020}
{Planck Collaboration}, {Aghanim}, N., {Akrami}, Y., {et~al.} 2020{\natexlab{b}}, \aap, 641, A5

\bibitem[{Platen {et~al.}(2007)Platen, Van De~Weygaert, \& Jones}]{Platen2007}
Platen, E., Van De~Weygaert, R., \& Jones, B. J.~T. 2007, Mon. Not. Roy. Astron. Soc., 380, 551

\bibitem[{Pollina {et~al.}(2019)Pollina, Hamaus, Paech, Dolag, Weller, Sánchez, Rykoff, Jain, Abbott, Allam, Avila, Bernstein, Bertin, Brooks, Burke, Carnero Rosell, Carrasco Kind, Carretero, Cunha, D’Andrea, da Costa, De Vicente, DePoy, Desai, Diehl, Doel, Evrard, Flaugher, Fosalba, Frieman, García-Bellido, Gerdes, Giannantonio, Gruen, Gschwend, Gutierrez, Hartley, Hollowood, Honscheid, Hoyle, James, Jeltema, Kuehn, Kuropatkin, Lima, March, Marshall, Melchior, Menanteau, Miquel, Plazas, Romer, Sanchez, Scarpine, Schindler, Schubnell, Sevilla-Noarbe, Smith, Soares-Santos, Sobreira, Suchyta, Tarle, Walker, Wester, \& Collaboration)}]{Pollina2019}
Pollina, G., Hamaus, N., Paech, K., {et~al.} 2019, Mon. Not. Roy. Astron. Soc., 487, 2836

\bibitem[{{Sachs} \& {Wolfe}(1967)}]{Sachs1967}
{Sachs}, R.~K. \& {Wolfe}, A.~M. 1967, \apj, 147, 73

\bibitem[{{Schlafly} {et~al.}(2019){Schlafly}, {Meisner}, \& {Green}}]{UnWISE}
{Schlafly}, E.~F., {Meisner}, A.~M., \& {Green}, G.~M. 2019, \apjs, 240, 30

\bibitem[{{Sutter} {et~al.}(2014){Sutter}, {Lavaux}, {Wandelt}, {Weinberg}, {Warren}, \& {Pisani}}]{Sutter2014}
{Sutter}, P.~M., {Lavaux}, G., {Wandelt}, B.~D., {et~al.} 2014, Mon. Not. Roy. Astron. Soc., 442, 3127

\bibitem[{{Tinker} {et~al.}(2008){Tinker}, {Kravtsov}, {Klypin}, {Abazajian}, {Warren}, {Yepes}, {Gottl{\"o}ber}, \& {Holz}}]{Tinker2008}
{Tinker}, J., {Kravtsov}, A.~V., {Klypin}, A., {et~al.} 2008, \apj, 688, 709

\bibitem[{{Totsuji} \& {Kihara}(1969)}]{Totsuji1969}
{Totsuji}, H. \& {Kihara}, T. 1969, \pasj, 21, 221

\bibitem[{Trümper(1993)}]{Truemper93}
Trümper, J. 1993, Science, 260, 1769

\bibitem[{{Wright} {et~al.}(2010){Wright}, {Eisenhardt}, {Mainzer}, {Ressler}, {Cutri}, {Jarrett}, {Kirkpatrick}, {Padgett}, {McMillan}, {Skrutskie}, {Stanford}, {Cohen}, {Walker}, {Mather}, {Leisawitz}, {Gautier}, {McLean}, {Benford}, {Lonsdale}, {Blain}, {Mendez}, {Irace}, {Duval}, {Liu}, {Royer}, {Heinrichsen}, {Howard}, {Shannon}, {Kendall}, {Walsh}, {Larsen}, {Cardon}, {Schick}, {Schwalm}, {Abid}, {Fabinsky}, {Naes}, \& {Tsai}}]{Wright_2010}
{Wright}, E.~L., {Eisenhardt}, P. R.~M., {Mainzer}, A.~K., {et~al.} 2010, \aj, 140, 1868

\bibitem[{{Zel'dovich}(1970)}]{Zel'dovich1970}
{Zel'dovich}, Y.~B. 1970, \aap, 5, 84

\bibitem[{{Zenteno} {et~al.}(2025){Zenteno}, {Kluge}, {Kharkrang}, {Hernandez-Lang}, {Damke}, {Saro}, {Monteiro-Oliveira}, {Carrasco}, {Salvato}, {Comparat}, {Fabricius}, {Snigula}, {Arevalo}, {Cuevas}, {Nilo Castellon}, {Ramirez}, {V{\'e}liz Astudillo}, {Landriau}, {Myers}, {Schlafly}, {Valdes}, {Weaver}, {Mohr}, {Grandis}, {Klein}, {Liu}, {Bulbul}, {Zhang}, {Sanders}, {Bahar}, {Ghirardini}, {Ramos}, \& {Balzer}}]{DEROSITAS}
{Zenteno}, A., {Kluge}, M., {Kharkrang}, R., {et~al.} 2025, \aap, 698, A171

\bibitem[{{Zheng} {et~al.}(2005){Zheng}, {Berlind}, {Weinberg}, {Benson}, {Baugh}, {Cole}, {Dav{\'e}}, {Frenk}, {Katz}, \& {Lacey}}]{Zheng2005}
{Zheng}, Z., {Berlind}, A.~A., {Weinberg}, D.~H., {et~al.} 2005, \apj, 633, 791

\bibitem[{{Zhou} {et~al.}(2023){Zhou}, {Ferraro}, {White}, {DeRose}, {Sailer}, {Aguilar}, {Ahlen}, {Bailey}, {Brooks}, {Claybaugh}, {Dawson}, {de la Macorra}, {Dey}, {Doel}, {Font-Ribera}, {Forero-Romero}, {Gontcho A Gontcho}, {Guy}, {Kremin}, {Lambert}, {Le Guillou}, {Levi}, {Magneville}, {Manera}, {Meisner}, {Miquel}, {Moustakas}, {Myers}, {Newman}, {Nie}, {Percival}, {Rezaie}, {Rossi}, {Sanchez}, {Schlegel}, {Schubnell}, {Seo}, {Tarl{\'e}}, \& {Zhou}}]{Zou2023}
{Zhou}, R., {Ferraro}, S., {White}, M., {et~al.} 2023, \jcap, 2023, 097

\end{thebibliography}

\appendix

\section{SNR calculations}\label{sec:SNR}

To quantify the SNR of the cross-correlations measured later in this work, we compute the $\chi^2$ statistic relative to a null hypothesis with zero signal and convert it to a z-score. This z-score directly represents the statistical significance of the detection in units of standard deviations of the data relative to the null. 

We note that simply taking $\sqrt{\chi^2}$ as the detection significance implicitly assumes a single degree of freedom. In our case, the cross-correlations have an unknown number of degrees of freedom, so the corresponding $\chi^2$ includes both noise and signal contributions. We therefore convert the $\chi^2$ into a p-value and then into a z-score, ensuring that correlations and the correct number of degrees of freedom are taken into account when evaluating the significance of the data. The $\chi^2$ is calculated as

\begin{equation}
\chi^2 = \mathbf{d}^\top \alpha \mathbf{C}^{-1} \mathbf{d} \, ,
\end{equation}
where $d$ is the data vector, $C$ is the corresponding covariance matrix, and $\alpha$ is the Hartlap factor \citep{Anderson2003}. The Hartlap factor accounts for the bias induced when inverting a noisy covariance matrix, which is given by

\begin{equation}
\alpha = \frac{N - p - 2}{N - 1} \, ,
\end{equation}
where $N$ is the number of samples used to estimate the covariance matrix, and $p$ is the number of bins with which the data vector is measured.

The covariance matrix is calculated by using the jackknife method on an object-by-object basis, where individual Cosmic Tunnels are removed to create jackknife samples. We note that generally it is more robust to use the jackknife method with patch-by-patch sampling, where whole patches of the sky are removed to create jackknife samples, which captures the spatial correlations of the objects. However, it has been shown that for 3D voids, the object-by-object jackknife agrees very well with the patch-by-patch jackknife, due to the large sizes of voids \cite{Demirbozan2024}. Furthermore, the Cosmic Tunnels studied here are significantly larger than typical 3D galaxy voids. We therefore expect the object-by-object jackknife to be robust in this case. The covariance matrix calculated here also accounts for the duplicate information present when the Cosmic Tunnel catalogue includes overlapping objects, thus ensuring that our SNR measurements are not overestimated. 

Next, we must estimate the number of degrees of freedom in a given cross-correlation. For the case of a perfectly Gaussian summary statistic with no bin-to-bin correlations (a diagonal covariance matrix), the number of bins would be equal to the number of degrees of freedom. However, that is not the case here, so we instead use the covariance matrix to estimate the number of degrees of freedom \citep{DelGiudice2021}. First, we obtain the eigenvalues of the covariance matrix by diagonalising it:
\begin{equation}
\mathbf{C} = \mathbf{E}\mathbf{\Lambda}\mathbf{E}^\top,
\end{equation}
where $\mathbf{\Lambda} = \mathrm{diag}(\lambda_1,\dots,\lambda_p)$, $\lambda_i$ is the $i^{th}$ eigenvalue and $\mathbf{E}$ is the orthonormal matrix of eigenvectors. These eigenvalues describe the variance along each principal component direction of the data. The total variance can then be given by summing all eigenvalues:
\begin{equation}
\lambda_{\mathrm{tot}} = \sum_{i=1}^{p} \lambda_i,
\end{equation}
each eigenvalue can therefore be normalised to represent the fractional contribution to the total variance
\begin{equation}
\overline{\lambda}_i = \frac{\lambda_i}{\lambda_{\mathrm{tot}}},
\end{equation}
which defines a probability distribution over the eigenmodes. The Shannon entropy of this distribution is then computed as:
\begin{equation}
H = -\sum_{i=1}^{p} \overline{\lambda}_i \ln \overline{\lambda}_i,
\end{equation}
which quantifies how evenly the variance is distributed across the eigenmodes. If all modes contribute equally, the entropy is maximal; if only a few dominate, the entropy is low. Finally, the effective number of degrees of freedom is given by the exponential of the Shannon entropy:
\begin{equation}
n_{\mathrm{eff}} = \exp(H).
\end{equation}
This value $n_{\mathrm{eff}}$ can be interpreted as the number of statistically independent components in the data vector.

Next, we compute the $p$-value associated with the measured $\chi^2$ and the effective number of degrees of freedom $n_{\mathrm{eff}}$. The p-value represents the probability of obtaining a $\chi^2$ value at least as large as the observed one under the null hypothesis, and is given by:

\begin{equation}
p = 1 - F_{\chi^2}(\chi^2\,|\,n_{\mathrm{eff}}),
\end{equation}
where $F_{\chi^2}(\chi^2\,|\,n_{\mathrm{eff}})$ is the cumulative distribution function of the chi-squared distribution with $n_{\mathrm{eff}}$ degrees of freedom, evaluated at the measured $\chi^2$. To express this in terms of a Gaussian-equivalent significance, we convert the p-value to a two-sided $Z$-score:

\begin{equation}
Z = \Phi^{-1}(1 - p/2),
\end{equation}
where $\Phi^{-1}$ is the inverse cumulative distribution function of the standard normal distribution. The quantity $Z$ represents the number of standard deviations by which the null hypothesis is disfavoured, which we use to quantify the statistical significance of the detections.

\end{document}